\documentclass[pra,twocolumn,showpacs,preprintnumbers,floatfix]{revtex4-1}
\usepackage{graphicx}
\usepackage{dcolumn}
\usepackage{bm}
\usepackage{latexsym,epsfig}
\usepackage{graphicx}
\usepackage{verbatim}
\usepackage[T1]{fontenc}
\usepackage{comment}
\usepackage{amsmath}
\usepackage{amssymb}
\usepackage{stmaryrd}
\usepackage{color,soul}
\usepackage{commath}

\usepackage{hyperref}
\hypersetup{
    colorlinks=true,      
    urlcolor=blue,
    citecolor=blue,
    linkcolor=blue
}
\usepackage{calc}
\newlength{\depthofsumsign}
\newlength{\totalheightofsumsign}
\newlength{\heightanddepthofargument}

\newcommand{\nsum}[1][1.4]{
    \mathop{%
        \raisebox
            {-#1\depthofsumsign+1\depthofsumsign}
            {\scalebox
                {#1}
                {$\displaystyle\sum$}%
            }
    }
}

\makeatletter
\newcommand*{\DivideLengths}[2]{%
  \strip@pt\dimexpr\number\numexpr\number\dimexpr#1\relax*65536/\number\dimexpr#2\relax\relax sp\relax
}
\makeatother

\newcommand{\beq}{\begin{equation}}
\newcommand{\eeq}{\end{equation}}
\newcommand{\bea}{\begin{eqnarray}}
\newcommand{\eea}{\end{eqnarray}}
\newcommand{\ben}{\begin{eqnarray*}}
\newcommand{\een}{\end{eqnarray*}}
\newcommand{\bfig}{\begin{figure}}
\newcommand{\efig}{\end{figure}}

\begin{document}
\title{Mott insulator phases of non-locally coupled bosons in bilayer optical superlattices}
\author{Sayan Lahiri$^{1}$, Suman Mondal$^1$, Manpreet Singh$^1$ and Tapan Mishra$^1$}
\affiliation{$^1$Department of Physics, Indian Institute of Technology, Guwahati-781039, India}
\date{\today}

\begin{abstract}
We investigate the ground state properties of a non-locally coupled bosonic system in a bilayer optical superlattice by considering bosons in one layer to be of softcore in nature and separately allowing 
two and three body hardcore constraints on the other layer. We find that the presence of different constraints on bosons in 
one layer influences the overall phase 
diagram exhibiting various Mott insulator phases at incommensurate densities due to the presence of the superlattice potential 
apart from the usual Mott insulators at commensurate densities. Moreover, the 
presence of two or three-body constraints significantly modifies the Mott insulator-Superfluid phase transition points as a function of the superlattice potential. 
Due to the various competing interactions, constraints and superlattice potential the phase diagrams exhibit significantly different features.  We 
obtain the complete phase diagrams by using the cluster-mean-field theory approach. We further extend this work 
to a coupled two-leg ladder superlattice where we obtain similar physics using the density matrix renormalization group method . 
\end{abstract}

\maketitle

\section{Introduction}
Quantum phase transitions in systems of ultracold quantum gases in optical lattices have revealed a host of new phenomena 
in recent years. The rapid progress in experimental manipulation of these weakly interacting systems in optical lattices has 
allowed to uncover various interesting physics which were impossible to understand using the conventional solid state systems. One such 
example is the observation of the superfluid (SF) to Mott insulator (MI) phase transition in a three-dimensional optical 
lattice~\cite{bloch}, the phenomenon which was predicted in the context of the Bose-Hubbard model~\cite{FisherPRB1989, jaksch}. 
The state-of-the-art experimental set ups and the flexibility to control the system parameters in such systems have provided a 
new platform to explore interesting phenomena in nature which has led to a plethora of novel and exciting physics. While 
the dominant interaction in such ultracold systems is the two-body contact interactions, it has been shown that there 
exist higher order local interactions as well, which have non-negligible effects on the ground state properties 
~\cite{wills, mark2011}. Recently it was shown that the leading multi-body interactions, such as the three-body 
interaction, can be engineered under suitable conditions in optical lattices~\cite{daley,tiesinga, petrov1, petrov2} which can play
important roles in discovering many-body
induced quantum phases, especially at higher densities~\cite{manpreetsuplat,sowinski1}. Such many-body interactions 
can drastically modify the behaviour of the system. For instance, the three-body interaction can become very large 
leading to the three-body hardcore constrained where not more than two particles can occupy a single site. This feature is crucial
in studying the systems of attractive bosons in optical lattices by preventing the collapse of bosons. These three-body constrained bosons(TBCs) are shown to
exhibit the superfluid to pair-superfluid(PSF) phase transition in optical lattice ~\cite{baranovprl}. Similar 
situation is also seen in the regime of large onsite two-body repulsion where no two bosons can simultaneously occupy a 
single site. In such a situation the bosons are called the hardcore bosons(HCBs) or the Tonk's gas~\cite{paredes1}. 

On the other hand long-range dipolar interactions in atoms, molecules and Rydberg atoms have culminated into a completely new realm of physics where 
several novel phenomena have been predicted and observed in recent experiments such as the CDW phases,the exotic supersolid phases and self-bound quantum droplets~\cite{baranovrev,
Ferlaino1, Ferlaino2, tanzi2019, bottcher2019}.
Interestingly, these non-local interactions have shown to couple systems which are spatially separated from 
each other such as the bilayer systems and two leg ladders. 
In such a scenario one can drive the two decoupled systems together with the help of the long-range dipole-dipole interaction~\cite{lewenstein,sansone,santos_arturo,manpreet}. Moreover, the bilayer systems
with inter-layer interactions resemble systems of two component atomic mixtures in optical lattices. 
A new area of research has evolved in the context of the dual species bosons, fermions as well as Bose-Fermi mixtures due to the 
recent advancement in cooling and trapping of binary mixtures in experiments~\cite{Inguscio,Weiman,catani1,trotzky1,jordens1,costa1,bloch2011,moritz1,roati1,best1}. 
The creation and manipulation of such dual-species mixtures with completely two different 
species of atoms or two hyperfine states of a particular species in optical lattices to acheive strong correlations have 
opened up various avenues in addressing complex many-body systems. 
The presence of different types of interactions 
compared to the single species systems have made the binary mixture a topic of great interest, as a result several theoretical predictions 
and experimental observations have been made in various 
context~\cite{altman,batrouni,mishra1,mishra2,ozaki1,pu1,kuklov1,mathey1,isacsson1,Sugawa2011,takahashi,moritz1,roati1,best1,costa1}.
At the same time the creation of optical superlattices~\cite{Porto1,Porto2} have proven to provide an additional 
flexibility to manipulate lattice potentials and periodicity which 
results in different interesting applications. A great deal of research has been done on optical superlattices and several new phases have been 
predicted in theory and observed in experiments on various context~\cite{manpreetsuperlat1,manpreetsuperlat2,aryasuperlat1,aryasuperlat2,aryasuperlat3,roth1,roth2,roux1,Porto1,sebby1,cheinet1,fleischhauer,Atala2013}.   
Although various investigations have been made in systems of ultracold quantum gases in optical superlattices, the study of bilayer superlattice 
or binary mixture in superlattices may lead to novel 
phenomena. As the systems of dual species mixture can be mapped to the spin systems under proper conditions, it promises a direct connection 
to the many-electron systems and magnetism~\cite{Altman_2003}. 
One such recent studies on multi-component bosons in optical superlattices has predicted various gapped phases using the single site mean-field 
approximation where the bosons are 
assumed to be softcore in nature and the intra-species interactions are considered to be of identical strength~\cite{shuchen1}. 
As a result the influence of one species of atoms on 
the other and vice-versa is similar in nature. 

\begin{figure}
\centering
\begin{minipage}[b]{.25\textwidth}
  \centering
  \includegraphics[width=0.8\linewidth]{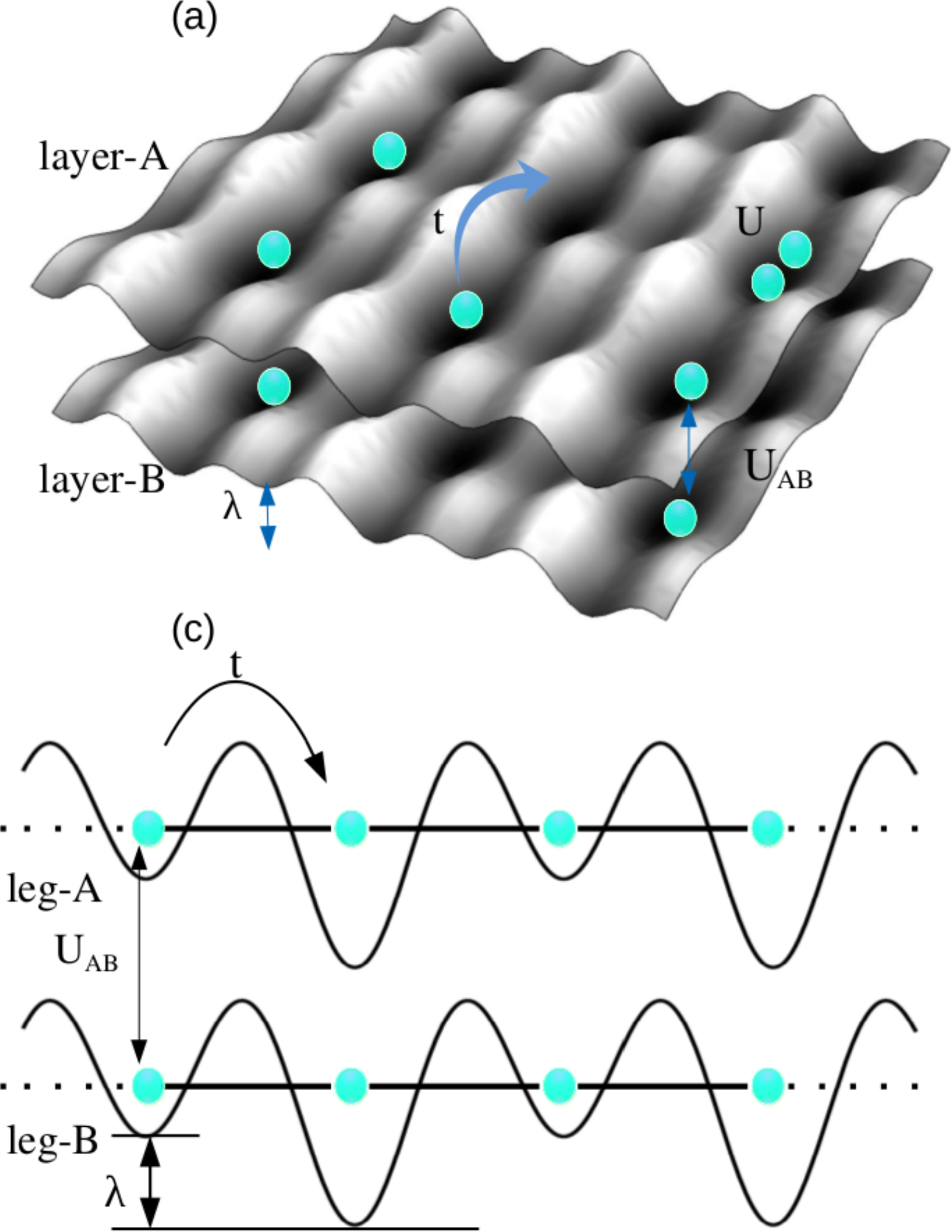}
\end{minipage}%
\begin{minipage}[b]{.25\textwidth}
  \centering
  \includegraphics[width=0.8\linewidth]{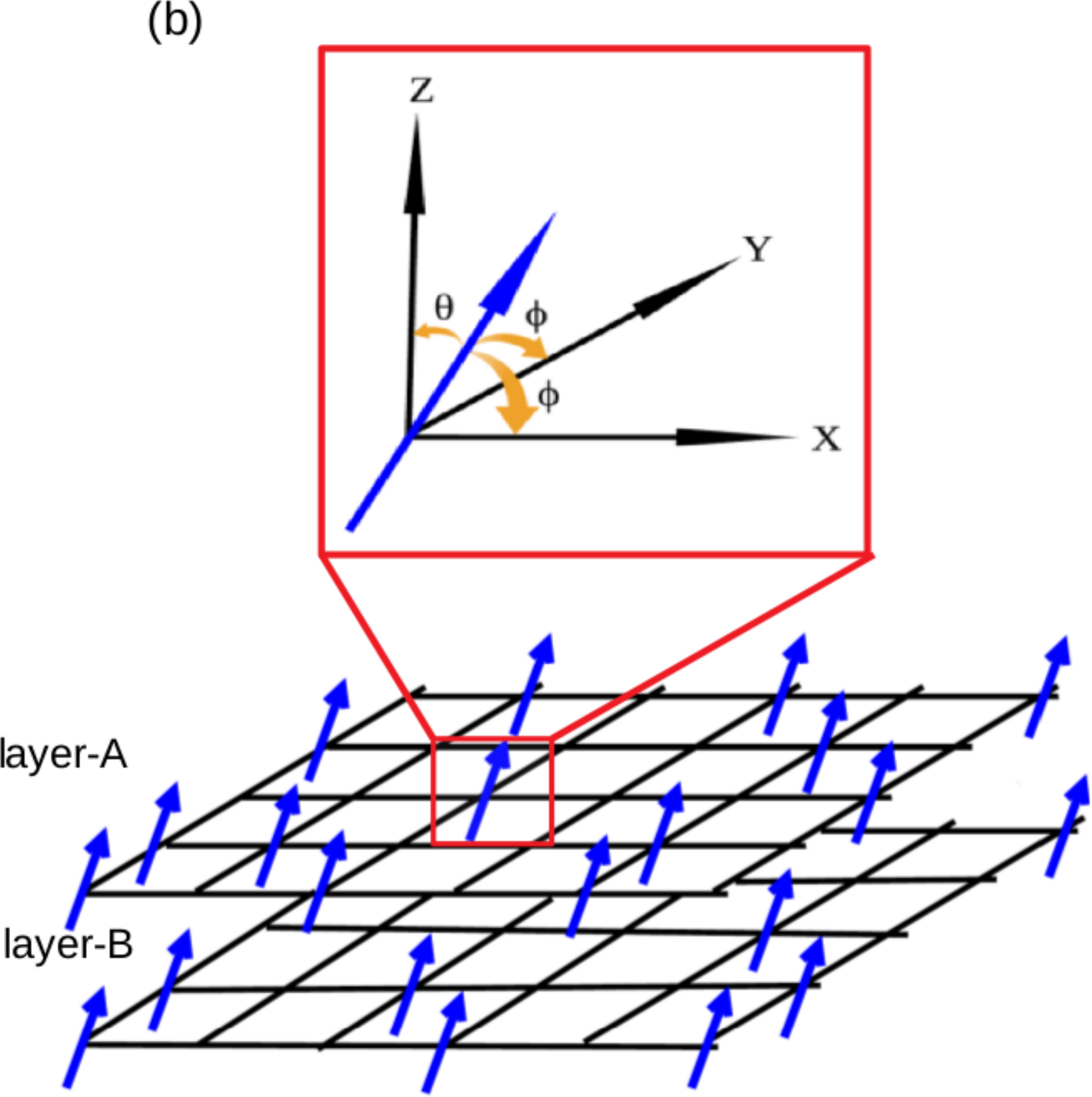}
\end{minipage}
\caption{(Color online)(a)Illustration of a bilayer superlattice having periodicity equal to two showing the intra-layer and inter-layer interactions.
(b) The possible alignment of the dipoles in this bilayer system. In the inset we have shown one dipole placed in the origin 
in such a way that it makes an angle $\phi$ with the $x$ and $y$ axes and $\theta$ with the $z$ axis. (c) Two leg optical superlattice
with leg-A and leg-B containing the softcore bosons and constrained bosons respectively.
These two legs interact among themselves via the non-local interaction $U_{AB}$.}
\label{fig:superlattice_with_period1}
\end{figure}

At this point we consider a bilayer bosonic system in optical superlattice as depicted in Fig.~\ref{fig:superlattice_with_period1}(a). The construction of the 
bilayer system is in such a way that both the layers are identical to 
each other in terms of lattice translation and a two-period superlattice potential is present only along the $x$-direction. 
As a result, the lattice periodicity is doubled in the $x$-direction where as 
along the $y$-direction there is no change in the periodicity. In principle, one can consider the superlattice potential along both the directions 
which is expected to exhibit similar particle dynamics in 
both the directions. However, the choice made in this work provides a situation where the particles tend to localise in every alternate sites due to 
the superlattice potential 
along one direction where as they are free to move on the other direction. In such a system we consider the bosons in 
layer-A to be of softcore in nature where as bosons in layer-B experience two or three-body hardcore constraint. 
We assume the particles are of dipolar in nature in both the layers and there exists 
only inter-layer interactions and the intra-layer interactions are 
suprressed. 
This situation can be achieved by orienting the dipoles in such a way that they are at 
magic angles with the line joining two nearest neighbour dipoles along both $x$- and $y$-directions of the layer. 
In such a scenario, the angle made with the line joining the dipoles sitting in two different layers are different from 
the magic angle resulting in a finite repulsive interaction which is proportional to $(1-3 cos^2\theta)$ as depicted in Fig.~\ref{fig:superlattice_with_period1}(b).

The model which describes such system is the modified Bose-Hubbard model given as:
\begin{eqnarray}
H=&-&t\nsum[1.]\limits_{\langle i,j \rangle,{\sigma\in[A,~B]}}({a}^\dagger_{i\sigma}{a}_{j\sigma} + H.c.)\nonumber\\
&+& \frac{U_{\sigma}}{2}\nsum[1.]\limits_{i,{\sigma\in[A,~B]}}\Bigg[{n}_{i\sigma}({n}_{i\sigma} - 1)-(\mu_{\sigma}-\lambda_i){n}_{i\sigma}\Bigg]\nonumber\\
&+&U_{AB}\sum\limits_i{n}_{iA}{n}_{iB}
\label{eqn:ham}
\end{eqnarray}
Here, ${a}^\dagger_{i\sigma}({a}_{i\sigma} )$ is the creation (annihilation) operator which creates (destroys) 
a boson in layer $\sigma(=A,B)$ and at site $i$, ${n}_{i\sigma} ={a}^\dagger_{i\sigma}{a}_{i\sigma}$ is the number 
operator and $t$ is the hopping amplitude between any two nearest neighbour sites $i$ and $j$.
While $U_{\sigma}$ represents the local two-body intra-layer interactions, $U_{AB}$ represents 
the inter-layer two body interaction.
$\mu_{\sigma}$ is the chemical potential and $\lambda_i$ is the superlattice potential along the x-direction which is 0($\lambda$) for odd(even) site indices as shown in 
Fig.~\ref{fig:superlattice_with_period1}(a).  The two and three body constraints in layer-B are achieved by considering $(a^{\dagger})^2=0$ and $(a^{\dagger})^3=0$ respectively. Note that for 
the HCBs in layer-B, $U_{\sigma=B}\rightarrow \infty$ and the terms associated to $U_{\sigma=B}$ will vanish in model(\ref{eqn:ham}) due to the hardcore constraint.  

The remaining part of the paper in organised as follows. In Sec. II, we give a brief discussion on the method used in 
this work. Section III contains the results and discussion of our work and we finally present the conclusion in Sec. IV.

\section{Method}
We investigate the ground state properties of the model given in Eq.~\ref{eqn:ham} by considering softcore bosons in layer-A whereas loading 
layer-B with HCBs and TBCs separately in two different scenarios. In both the cases we study the phase diagram of the system and obtain various 
gapped phases which appear at commensurate as well 
as incommensurate densities. We also show how the tip positions of these gapped phases or in other words the gapped to gapless phase transitions 
behave by changing the constraint on the bosons in layer-B. 
To this end we implement the cluster mean-field theory(CMFT) approach to analyse the Eq.~\ref{eqn:ham}. In the end we utilise the well known 
density matrix renormalization group (DMRG) method for the two-leg ladder model to examine the features in one dimension. 
It is to be noted that this problem can also be analysed using the simple mean-field 
decoupling approximation~\cite{sheshadri1} which can capture the qualitatve physics of the system. 
However, in order to achieve better accuracy we employ the CMFT approach. For the models like shown in Eq.~\ref{eqn:ham}, 
the CMFT method works fairly well with less computational 
complexities and may approach the Quantum Monte Carlo results in the thermodynamic limit for some specific situations~\cite{manpreet,danshita8,chen2017two}. 

In the CMFT method the entire system is divided into identical clusters of limited number of sites which can be treated exactly 
and then the coupling between different clusters are 
treated in a mean-field way. The accuracy of this method improves by increasing the number of sites in the cluster. 
With this approximation the original Hamiltonian of Eq.~\ref{eqn:ham} can be written as 
\begin{equation}
 {H}={H}_{C}+{H}_{MF}
 \label{eqn:4}
\end{equation}
where $H_C(H_{MF})$ is the cluster(mean-field) part of the Hamiltonian. $H_C$ is same as Eq.~\ref{eqn:ham} limited to the cluster size. The hoppings between the clusters are 
then approximated using the simple mean-field type decoupling scheme given as 
\begin{eqnarray}
 {a}^\dagger_{i\sigma}{a}_{j\sigma}={\langle{a}^\dagger_{i\sigma}\rangle}{a}_{j\sigma}+{{a}^\dagger_{i\sigma}}\langle{a}_{j\sigma}\rangle
 -{\langle{a}^\dagger_{i\sigma}\rangle}{\langle{a}_{j\sigma}\rangle}
\label{eqn:3}
\end{eqnarray}
Introducing the two layer-dependent 
SF order parameter $\psi_{i\sigma}={\langle{a}^\dagger_{i\sigma}\rangle}={\langle{a}_{i\sigma}\rangle}$  and using Eq. \ref{eqn:3} we write the ${H}_{MF}$ as 
\begin{eqnarray}
 {H}_{MF} &=& -t\sum\limits_{\sigma,\langle i,j \rangle}[(a^\dagger_{i\sigma}+a_{i\sigma})\psi_{j\sigma}-\psi_{i\sigma}^* \psi_{j\sigma}]
\end{eqnarray}

In our calculation we have set $t = 1$ to make all the physical quantities dimensionless. For the CMFT calculation we use a four sites cluster which consists
of two sites from both the layers. We call this a supercell in the following discussion. We
define the quantity $n = \sum\limits_{\sigma \in[A,B]}\sum\limits_{i=1}^{2}n_{i\sigma}$ and $\rho=\frac{1}{4}n$ which are the total particle number
and density in one supercell to distinguish various 
phases. 
We also assume equal chemical potentials for bosons in both the layers by making $\mu_A=\mu_B=\mu$ in our calculation. By 
fixing the values of $U_A$, $U_B$(in case of TBCs) and $U_{AB}$ we compute the complete phase diagram in the $\mu$ vs $\lambda$ plane. 
One can in principle fix the value of $\lambda$ and vary the 
interactions as done in Ref.~\cite{shuchen1}. However, to understand the effect of the superlattice potential on the phase diagram, 
we consider a range of $\lambda$ in our calculation.
Further we analyse the system in one dimension by considering a two-leg ladder which can be viewed as two superlattice
chains coupled via the dipole-dipole interaction. We study 
the model(\ref{eqn:ham}) by using the DMRG method by considering $\sigma$ as the two legs of the ladder. 
In our DMRG calculation we assume system sizes up to $100$ sites and $500$ density 
matrix eigenstates. The cluster and system size considered in our calculation are found to be sufficent to capture the physics we are interested in.

\section{Results and Discussion}
\subsection{Hardcore constraint for bosons in layer-B}
\begin{figure}[b]
  \centering
  \includegraphics[width=1\linewidth]{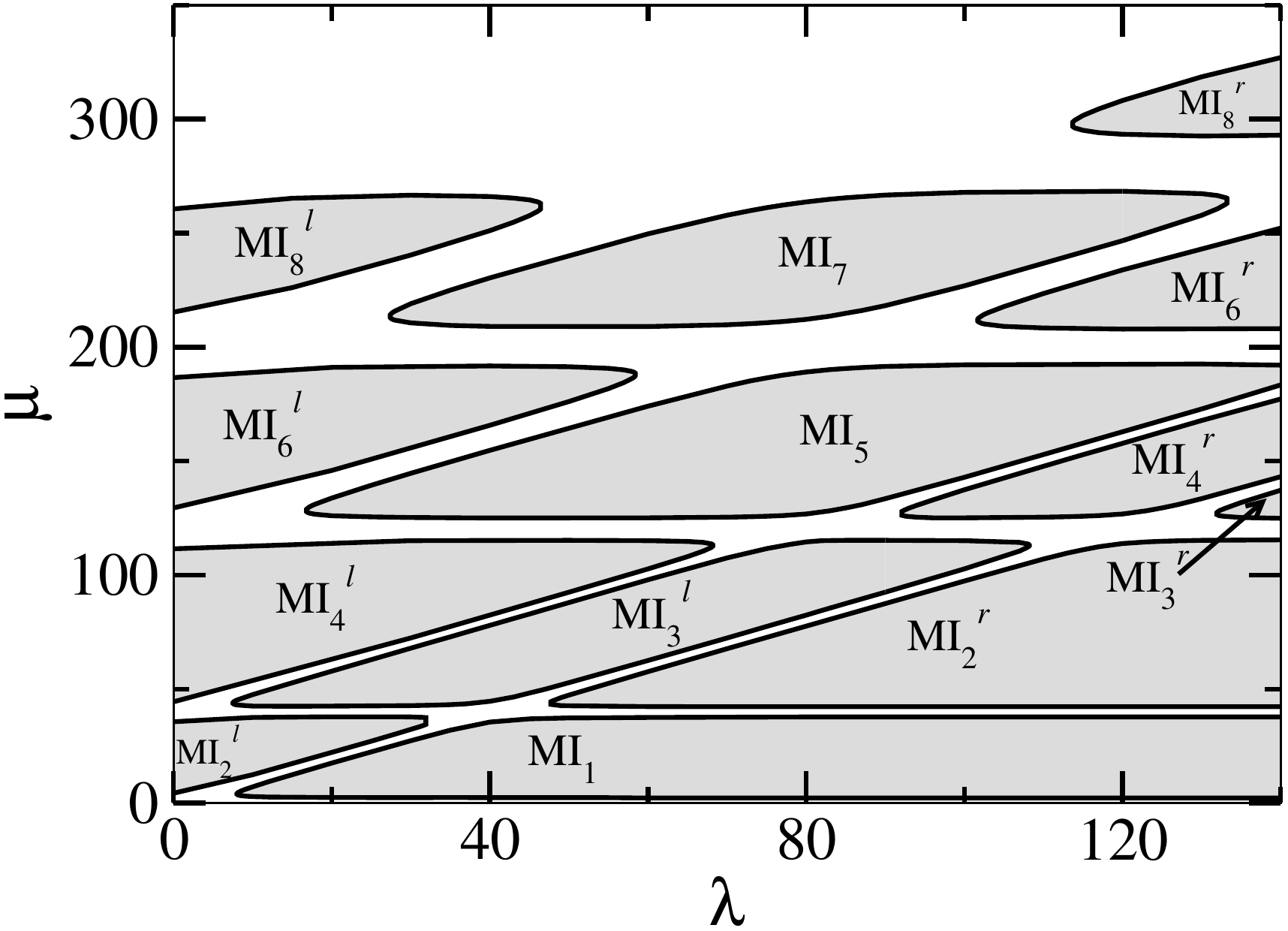}
  \caption{CMFT phase diagram when bosons in layer-A are softcore and in layer-B are HCBs. Here $U_A=80$ and $U_{AB}=40$.}
\label{fig:cmftphasedia1}
\end{figure}
In this part, we discuss the case when the bosons in layer-B are hardcore in nature.
In the decoupled layer limit i.e. $U_{AB}=0$, both the layers behave as independent two dimensional ($2d$) systems. 
It is well known that there exists a critical value of $\lambda$ for which the system undegoes an SF to MI transitions at half filling for 
both hardcore and softcore bosons in $2d$~\cite{PhysRevA.96.011602,manpreetsuplat,roth1}. 
Therefore, it is expected that the first Mott lobe would appear in the system after a critical value of $\lambda$ for density $\rho=1/4$ where any one 
of the layers attains half filling. 
At this stage the 
inter-layer interaction $U_{AB}$ has no role on the phase diagram as the particles reside in any one of the layer. 
However, with the increase in the chemical potential $\mu$ both the 
layers get populated and 
one may see interesting interplay between $U_A$, $U_{AB}$ and $\lambda$ which leads to various gapped phases and transitions to 
the SF phase at different integer and non integer fillings of individual 
layers as shown in  Fig.~\ref{fig:cmftphasedia1}. 
In this paper we call all the gapped phases the MI phases although 
the ones at non integer densities are different from the usual MI phase where each site is occupied 
with same integer number of bosons~\cite{manpreetsuperlat2}. However, in the case of superlattices, one 
can consider the density with respect to the unit-cell (the periodicity of the superlattice) so that the gapped phases 
at non integer densities can be called as the MI phases for those particular densities. 

\begin{figure}[b]
   \centering
\includegraphics[width=1.\linewidth]{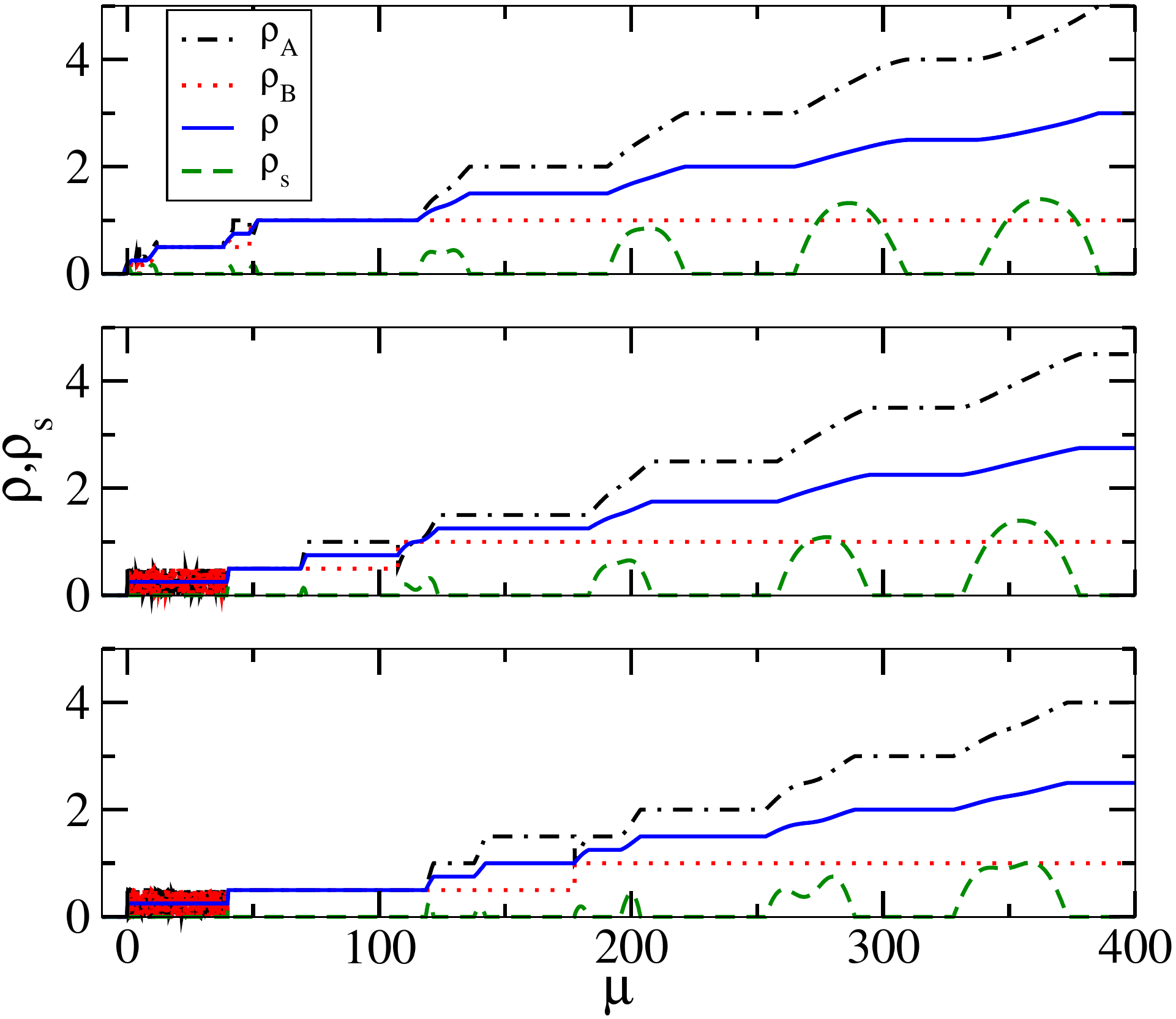}
\caption{(Color online)Variation of $\rho$(solid blue curve) and $\rho_s$(dashed green curve) with respect to $\mu$ for $\lambda=10,~70,~140$ shows the gapped and gapless phases when layer-B has HCBs. 
The plateaus in $\rho$ correspond to the gap in the MI phases whereas the 
shoulders around the plateaus (where the values of $\rho_s$ are finite) indicate the gapless SF phase. We also plot the individual layer densities as $\rho_A$(dot dashed curve) and $\rho_B$(dotted curve) 
for clarity. Due to the hardcore nature $\rho_B$ saturates at one. The fluctuations in $\rho_A$ and $\rho_B$ are due to the degenerate states in the CMFT calculation.}
\label{fig:rhomu}
\end{figure}

The phase diagram of Fig.~\ref{fig:cmftphasedia1} is obtained by self consistently diagonalising the 
Hamiltonian shown in Eq.~\ref{eqn:4} to obtain the ground state wave function and then the superfulid order parameter $\psi$ as 
discussed in the previous section. 
By considering a large onsite interaction $U_A=80$ and $U_{AB}=U_A/2$
which is sufficient to establish various gapped phases for the softcore bosons in one layer and by varying $\lambda$ for a wide range of values, we obtain the 
entire phase diagram which consists of the gapped MI lobes and the intermediate SF phases. 
Here we define the total superfluid order parameter $\rho_s=\frac{1}{4}\sum_{i,\sigma}|\psi_{i\sigma}|^2$, where $\psi_{i\sigma}$ is the 
layer dependent superfluid order parameter as discussed in the previous section. The gapped phases are obtained by looking at the 
vanishing up of the total superfluid order parameter with resepect to the chemical potential in 
the $\rho_s$ vs $\mu$ plot for several values of $\lambda$. In Fig.~\ref{fig:rhomu} we show 
$\rho$(solid blue line) and $\rho_s$(dashed green line) with respect to $\mu$ for $\lambda =10,~70,~140$ which cut through the gapped phases in Fig.~\ref{fig:cmftphasedia1} 
parallel to the $\mu$ axis. It can be seen that for a particular $\lambda$, 
as $\mu$ increases the plateaus in $\rho$ appear and at the same time $\rho_s$ vanishes corresponding to the MI phases.
We denote the MI phases as MI$_n$ where the subscript $n$ indicates the 
total number of particles i.e. $n_A+n_B$ in an supercell. The possible supercell atom distribution for all the MI phases are presented in Table. I. 

\begin{table}
 \caption{Table depicting various MI states when atoms in layer-B are HCBs. Each state shows the density distribution in the supercell corresponding to a particular MI state for a given $n$ and $\rho$.}
 \begin{center}
 \begin{tabular}[c]{ |p{1cm}|p{1.5cm}|p{4cm}|}
 \hline
 \multicolumn{3}{|c|}{Boson distribution in the unit cell } \\
 \hline
n & $\rho$ & Supercell Configuration\\
 \hline
 \vspace{-0.05cm}$1$&\vspace{-0.05cm}$0.25$ &\vspace{-0.15cm} \resizebox{.2\textwidth}{!}{MI$_1=$ $\biggm\lvert\begin{array}{cccccc}
0 & 1    \\
0 & 0
\end{array}\biggm\rangle$~or~MI$_1=$ $\biggm\lvert\begin{array}{cccccc}
0 & 0    \\
0 & 1
\end{array}\biggm\rangle$}\vspace{0.05cm}\\ 
\hline
\vspace{-0.05cm}$2$&\vspace{-0.05cm}$0.5$ &\vspace{-0.15cm} \resizebox{.2\textwidth}{!}{MI$_2^l=$ $\biggm\lvert\begin{array}{cccccc}
0 & 1    \\
1 & 0
\end{array}\biggm\rangle$~or~MI$_2^l=$ $\biggm\lvert\begin{array}{cccccc}
1 & 0    \\
0 & 1
\end{array}\biggm\rangle$} \newline \resizebox{.1\textwidth}{!}{and MI$_2^r=$ $\biggm\lvert\begin{array}{cccccc}
0 & 1    \\
0 & 1
\end{array}\biggm\rangle$}\vspace{0.05cm}\\
\hline
\vspace{-0.05cm}$3$&\vspace{-0.05cm}$0.75$ &\vspace{-0.15cm}\resizebox{.2\textwidth}{!}{MI$_3^l=$ $\biggm\lvert\begin{array}{cccccc}
1 & 1    \\
0 & 1
\end{array}\biggm\rangle$ and MI$_3^r=$ $\biggm\lvert\begin{array}{cccccc}
0 & 2    \\
0 & 1
\end{array}\biggm\rangle$}\vspace{0.05cm}\\
\hline
 \vspace{-0.05cm}$4$&\vspace{-0.05cm}$1.0$ &\vspace{-0.15cm}\resizebox{.2\textwidth}{!}{MI$_4^l=$ $\biggm\lvert\begin{array}{cccccc}
1 & 1    \\
1 & 1
\end{array}\biggm\rangle$ and MI$_4^r=$ $\biggm\lvert\begin{array}{cccccc}
1 & 2    \\
0 & 1
\end{array}\biggm\rangle$}\vspace{0.05cm}\\
\hline
\vspace{-0.05cm}$5$&\vspace{-0.05cm}$1.25$ &\vspace{-0.15cm}\resizebox{.1\textwidth}{!}{MI$_5=$ $\biggm\lvert\begin{array}{cccccc}
1 & 2    \\
1 & 1
\end{array}\biggm\rangle$}\vspace{0.05cm}\\ 
\hline
 \vspace{-0.05cm}$6$&\vspace{-0.05cm}$1.5$ &\vspace{-0.15cm}\resizebox{.2\textwidth}{!}{MI$_6^l=$ $\biggm\lvert\begin{array}{cccccc}
2 & 2    \\
1 & 1
\end{array}\biggm\rangle$ and MI$_6^r=$ $\biggm\lvert\begin{array}{cccccc}
1 & 3    \\
1 & 1
\end{array}\biggm\rangle$}\vspace{0.05cm}\\
\hline
\vspace{-0.05cm}$7$&\vspace{-0.05cm}$1.75$ &\vspace{-0.15cm}\resizebox{.1\textwidth}{!}{MI$_7=$ $\biggm\lvert\begin{array}{cccccc}
2 & 3    \\
1 & 1
\end{array}\biggm\rangle$}\vspace{0.05cm}\\ 
\hline
 \vspace{-0.05cm}$8$&\vspace{-0.05cm}$2.0$ &\vspace{-0.15cm}\resizebox{.2\textwidth}{!}{MI$_8^l=$ $\biggm\lvert\begin{array}{cccccc}
3 & 3    \\
1 & 1
\end{array}\biggm\rangle$ and MI$_8^r=$ $\biggm\lvert\begin{array}{cccccc}
2 & 4    \\
1 & 1
\end{array}\biggm\rangle$}\vspace{0.05cm}\\
\hline
\end{tabular}
\end{center}
\end{table}

It can be seen from Fig.~\ref{fig:cmftphasedia1} that the first lobe which corresponds to the MI$_1$ appears after a critical $\lambda\sim 8$ at $\rho=1/4$ as 
discussed before. The gap continues to be finite and 
the lobe expands as $\lambda$ increases further. Upon increasing the value of $\mu$ or in other words by increasing the particle number in the 
system, the other gapped phases start to appear, which are seen 
as plateaus in the $\rho$ vs $\mu$ plot shown in Fig. ~\ref{fig:rhomu}. For total density $\rho=1/2$(half filling of both the layers), 
there exists two gapped lobes separated by the SF phase as a function of $\lambda$. The appearance of large gap at vanishing $\lambda$ can be 
attributed to the effect of $U_{AB}$ which prevents the atoms in layer-A and layer-B to occupy the same sites. 
Hence, one may expect a gapped phase which is similar to the checker-board solid for the Bose-Fermi mixture on a square lattice~\cite{santosfermibose}.
We call this phase as the MI$_2^l$ phase. As discussed in Ref.~\cite{santosfermibose}, the 
stability of this gapped phase depends on the ratio $U_{AB}/U_A$. In our case $U_{AB}/U_A=0.5$, which is sufficient to open a gap in the system. However, the gap gradually decreases as the value 
of $\lambda$ increases and as a result the system enters into the SF phase. 
This is because of the increase in the effective onsite potential in every alternate sites on both the layers which results in a smaller 
ratio $U_{AB}/U_A$. Increasing in value of $\lambda$ further, the gapped phase reappears after a critical $\lambda\sim47.6$. 
 At this stage, the superlattice potential is very strong compared to the ratio $U_{AB}/U_A$ and the bosons reside in the 
deep lattice sites. We call this gapped phase the MI$_2^r$ phase which is similar to the striped phase for the $2d$ case. 
The density distribution can be seen from the $\rho$ vs. $\mu$ plot as shown in Fig.~\ref{fig:rhomu}(see figure for detail).

At this stage further increase in density results in the next gapped phases at $\rho=3/4$. The situation at this density is 
completely different from the case of half density. Here we find that the system is initially in the gapless SF phase for 
a range of $\lambda$ starting from $\lambda=0$ and there exists a gapped island for some intermediate range of $\lambda$ and then a gapped phase for the 
large values of $\lambda$. 
The physics at this density can be understood by the following analysis. For $\lambda=0$, as $\mu$ increases the layer-A will 
start to get populated first due to the 
softcore nature and all the sites are occupied by one atom each giving rise to unit filling and the layer-B remains at half filling. At 
unit filling the layer-A is in the MI phase as $U_A$ is sufficiently strong. As a result, the atoms 
in layer-B will experience equal repulsion $U_{AB}$ from all the sites of layer-A and 
hence they can move freely giving rise to the SF phase of layer-B. Therefore, the system as a whole is gapless although layer-A is in the MI state. 
The increase in $\lambda$, however, introduces 
the gap in the system by localising the hardcore bosons into the deep sites of layer-B while the layer-A remains in the MI phase. 
The resulting system is therefore, 
a gapped phase. This phase is called the MI$_3^l$ phase which lies between $\lambda\sim7.5$ and $\lambda\sim108.4$. Further increase in $\lambda$ leads to increase in particle-hole fluctuation and 
the gapped MI phase starts to melt and the SF phase reappears 
in the system. Eventually, the system enters into another gapped phase after a critical value of $\lambda\sim131.10$ where in layer-A two bosons are localized in the 
deep lattice sites. We call this phase as the MI$_3^r$ phase as depicted in the phase diagram of Fig.~\ref{fig:cmftphasedia1}.

At this stage, further increase in the value of $\mu$ will facilitate addition of bosons in layer-B which saturates at unit filling due to the hardcore nature of bosons  
where all the sites are occupied by one boson each. This situation corresponds to the MI$_4$ phases in the phase diagram where the 
total density of the system is $\rho=1$. When $\lambda$ is small, we have every sites in both the layers occupied by one atom each and 
this phase is called the MI$_4^l$. When $\lambda$ increases, the MI$_4^l$ phase melts and the system enters into the SF phase and eventually 
leading to the MI$_4^r$ phase where the atoms in layer-A occupy the deep sites while the layer-B maintains the uniform density due to the 
hardcore nature. In such a situation the layer-A is like the striped phase and layer-B is saturated. These features can be clearly seen from the individual layer densities as shown in Fig.~\ref{fig:rhomu}.

Similar situation arises for the 
other integer and non-integer densities where two distinct gapped phases appear at two limits of the superlattice potential which are separated by 
the SF phase as depicted in the phase diagram of Fig.~\ref{fig:cmftphasedia1}. The corresponding 
boson distributions are shown in Table.I for clarity. For quarter-integer densities, the gapped islands appear for a range of intermediate 
values of $\lambda$ separated by the SF phase. 
It is to be noted that the tip of the right lobes shifts towards larger values of $\lambda$ as $\rho$ increases. However, 
there occurs an interesting pattern for the left lobes where the tip first shifts towards larger $\lambda$ up to MI$_4^l$ and then shifts left for 
higher densities. The appearance of this feature is attributed to the presence of bosons in layer-B in all the lattice sites after a 
critical density $\rho\geq 1$.
At these densities, the bosons in layer-A does not get affected by the presence of the bosons in layer-B as it experiences uniform repulsion 
which is equal to 
$U_{AB}$ from all sites. Therefore, the physics of the system is governed only by the properties of bosons in layer-A as 
discussed in Ref.~\cite{manpreetsuplat}. 
The MI$^l$-SF transitions happens for smaller and smaller values of $\lambda$ as the density increases because the increase in density leads to the 
decrease in the effective onsite interactions on the shallow lattice sites. 
Therefore, the MI$^l$ lobes melt into the SF phases due to the hopping $t$ which dominates over the interactions.  
On the other hand the SF-MI$^r$ transition points shift towards the larger values of $\lambda$ at higher densities 
because of the increase in number of particles in the deep wells which results in an increase in $U_{AB}$. 
Therefore, a stronger $\lambda$ is necessary to introduce the MI$^r$ phases as 
can be seen from the phase diagram.  

\subsection{Three-body constraint for bosons in layer-B}

\begin{figure}[b]
  \centering
  \includegraphics[width=1\linewidth]{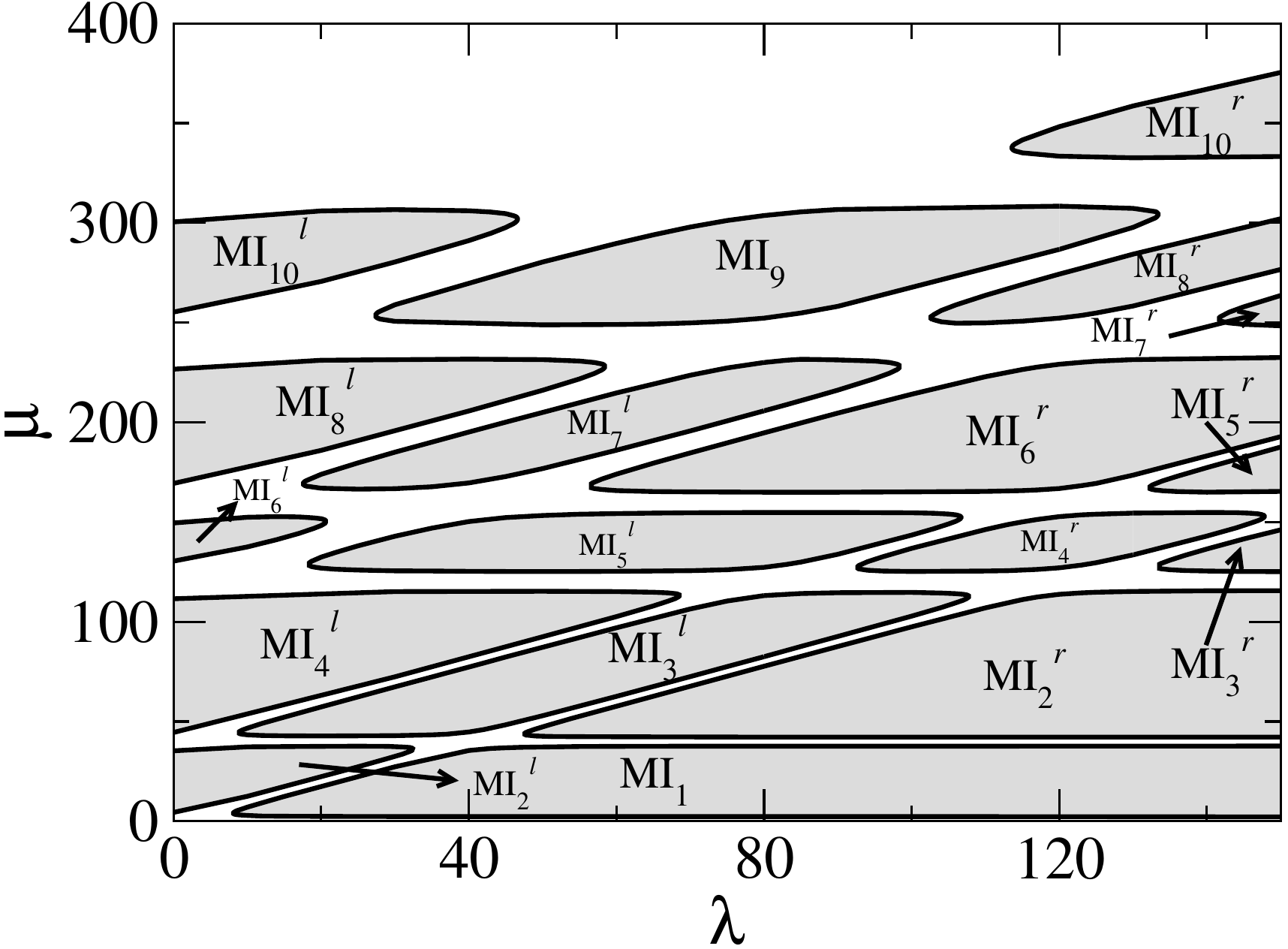}
  \caption{CMFT phase diagram when bosons in layer-A are softcore and in layer-B are TBCs. Here $U_A=U_B=80$ and $U_{AB}=40$.}
\label{fig:cmftphasedia2}
\end{figure}

\begin{figure}[b]
   \centering
\includegraphics[width=1.\linewidth]{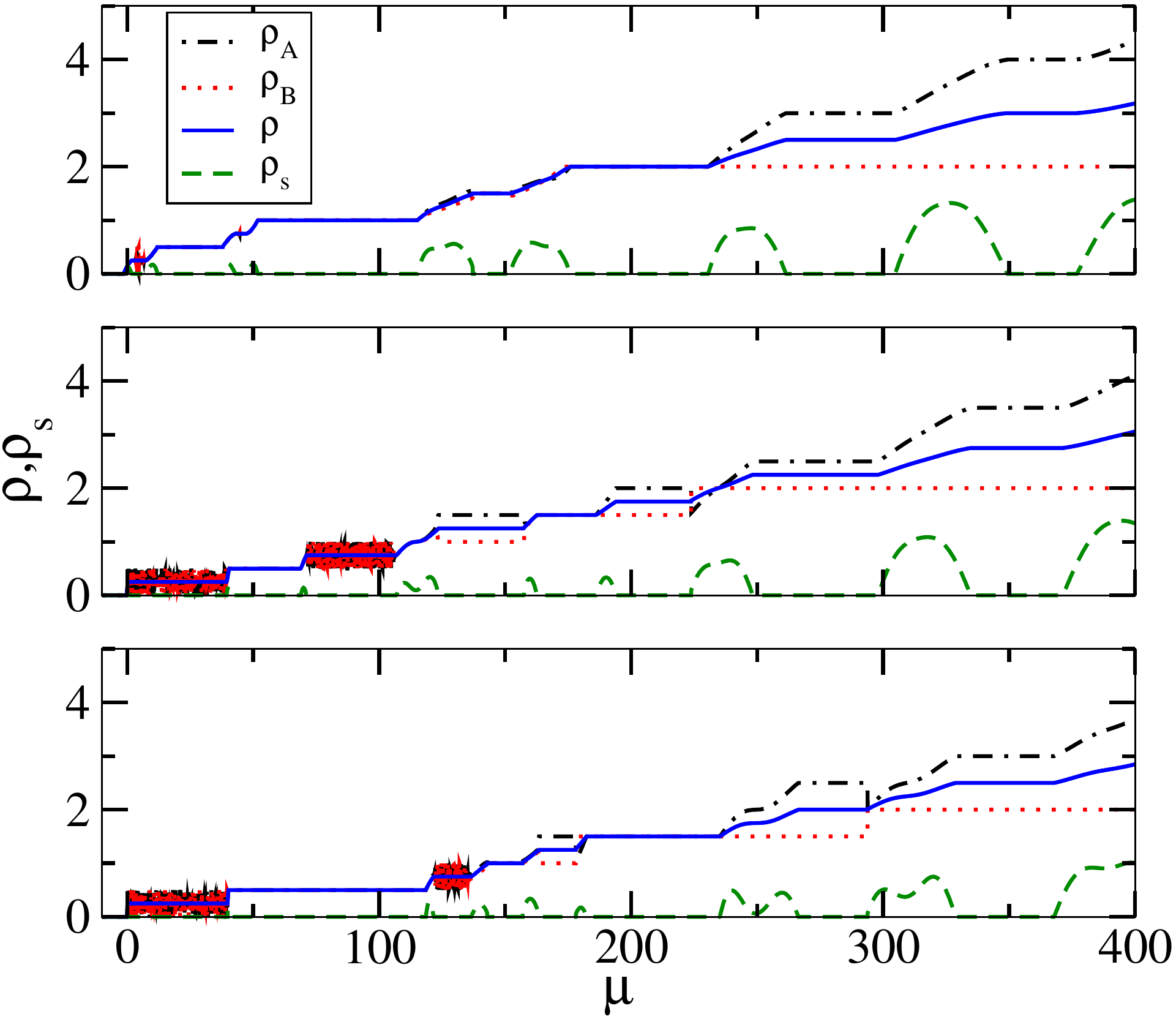}
\caption{(Color online)Figure shows the value of $\rho$(solid blue curve) and $\rho_s$(dashed green curve) with respect to $\mu$ for $\lambda=10,~70,~140$ corresponding to the gapped and gapless 
phases when bosons in layer-B are TBCs. 
The plateaus in $\rho$ indicates the gap in the MI phases whereas the 
shoulder around the plateaus (where the values of $\rho_s$ are finite) correspond to the gapless SF phase. 
We also plot the individual layer densities as $\rho_A$(dot dashed curve) and $\rho_B$(dotted curve) 
for clarity. Note that $\rho_B$ saturates at two after a certain chemical potential due to the three-body constraint.}
\label{fig:rhomu_tbc}
\end{figure}

Now we discuss the case in which we replace the bosons in layer-B with the TBCs. As discussed before, 
the effect of three-body constraint is a result of large three-body onsite repulsion i.e. $W\to\infty$. 
In such a situation the maximum number of bosons allowed per lattice 
site is two i.e. $(a^\dagger)^3=0$. Such constrained bosons may impart significant effects on the overall phase diagram of bosons in optical superlattice. 
Note that, in this case we have finite values of two-body interaction $U_B$ in layer-B. 
similar to the previous section, we numerically solve the mean-field Hamiltonian given in Eq.~\ref{eqn:4} and 
with $U_A=U_B=80$ and $U_{AB}=40$ and varying $\lambda$ we obtain various gapped phases.
The ground state phase diagram is shown in Fig.~\ref{fig:cmftphasedia2} and the particle distributions are shown in Table-II.  

\begin{table}
 \caption{Table depicting various MI states when atoms in layer-B are TBCs. Each state shows the density distribution in the supercell corresponding to a particular MI state for a given $n$ and $\rho$.}
 \begin{center}
 \begin{tabular}[c]{ |p{1cm}|p{1.5cm}|p{4cm}|}
 \hline
 \multicolumn{3}{|c|}{Boson distribution in the unit cell } \\
 \hline
n & $\rho$ & Supercell Configuration\\
 \hline
 \vspace{-0.05cm}$1$&\vspace{-0.05cm}$0.25$ &\vspace{-0.15cm} \resizebox{.2\textwidth}{!}{MI$_1=$ $\biggm\lvert\begin{array}{cccccc}
0 & 1    \\
0 & 0
\end{array}\biggm\rangle$~or~MI$_1=$ $\biggm\lvert\begin{array}{cccccc}
0 & 0    \\
0 & 1
\end{array}\biggm\rangle$}\vspace{0.05cm}\\ 
\hline
\vspace{-0.05cm}$2$&\vspace{-0.05cm}$0.5$ &\vspace{-0.15cm} \resizebox{.2\textwidth}{!}{MI$_2^l=$ $\biggm\lvert\begin{array}{cccccc}
0 & 1    \\
1 & 0
\end{array}\biggm\rangle$~or~MI$_2^l=$ $\biggm\lvert\begin{array}{cccccc}
1 & 0    \\
0 & 1
\end{array}\biggm\rangle$} \newline \resizebox{.1\textwidth}{!}{and MI$_2^r=$ $\biggm\lvert\begin{array}{cccccc}
0 & 1    \\
0 & 1
\end{array}\biggm\rangle$}\vspace{0.05cm}\\
\hline
\vspace{-0.05cm}$3$&\vspace{-0.05cm}$0.75$ &\vspace{-0.15cm}\resizebox{.2\textwidth}{!}{MI$_3^l=$ $\biggm\lvert\begin{array}{cccccc}
1 & 1    \\
0 & 1
\end{array}\biggm\rangle$ or MI$_3^l=$ $\biggm\lvert\begin{array}{cccccc}
0 & 1    \\
1 & 1
\end{array}\biggm\rangle$}
\newline \resizebox{.2\textwidth}{!}{and MI$_3^r=$ $\biggm\lvert\begin{array}{cccccc}
0 & 2    \\
0 & 1
\end{array}\biggm\rangle$ or MI$_3^r=$ $\biggm\lvert\begin{array}{cccccc}
0 & 1    \\
0 & 2
\end{array}\biggm\rangle$}\vspace{0.05cm}\\
\hline
 \vspace{-0.05cm}$4$&\vspace{-0.05cm}$1.0$ &\vspace{-0.15cm}\resizebox{.2\textwidth}{!}{MI$_4^l=$ $\biggm\lvert\begin{array}{cccccc}
1 & 1    \\
1 & 1
\end{array}\biggm\rangle$ and MI$_4^r=$ $\biggm\lvert\begin{array}{cccccc}
1 & 1    \\
0 & 2
\end{array}\biggm\rangle$} \newline \resizebox{.1\textwidth}{!}{or MI$_4^r=$ $\biggm\lvert\begin{array}{cccccc}
0 & 2    \\
1 & 1
\end{array}\biggm\rangle$}\vspace{0.05cm} \\
\hline
\vspace{-0.05cm}$5$&\vspace{-0.05cm}$1.25$ &\vspace{-0.15cm}\resizebox{.2\textwidth}{!}{MI$_5^l=$ $\biggm\lvert\begin{array}{cccccc}
1 & 2    \\
1 & 1
\end{array}\biggm\rangle$ and MI$_5^r=$ $\biggm\lvert\begin{array}{cccccc}
1 & 2    \\
0 & 2
\end{array}\biggm\rangle$}\vspace{0.05cm}\\ 
\hline
 \vspace{-0.05cm}$6$&\vspace{-0.05cm}$1.5$ &\vspace{-0.15cm}\resizebox{.2\textwidth}{!}{MI$_6^l=$ $\biggm\lvert\begin{array}{cccccc}
2 & 1    \\
2 & 1
\end{array}\biggm\rangle$ or MI$_6^l=$ $\biggm\lvert\begin{array}{cccccc}
1 & 2    \\
1 & 2
\end{array}\biggm\rangle$} \newline \resizebox{.2\textwidth}{!}{MI$_6^l=$ $\biggm\lvert\begin{array}{cccccc}
2 & 1    \\
1 & 2
\end{array}\biggm\rangle$ or MI$_6^l=$ $\biggm\lvert\begin{array}{cccccc}
1 & 2    \\
2 & 1
\end{array}\biggm\rangle$}
\newline \resizebox{.1\textwidth}{!}{and MI$_6^r=$ $\biggm\lvert\begin{array}{cccccc}
1 & 2    \\
1 & 2
\end{array}\biggm\rangle$}\vspace{0.05cm}\\
\hline
\vspace{-0.05cm}$7$&\vspace{-0.05cm}$1.75$ &\vspace{-0.15cm}\resizebox{.2\textwidth}{!}{MI$_7^l=$ $\biggm\lvert\begin{array}{cccccc}
2 & 2    \\
1 & 2
\end{array}\biggm\rangle$ and MI$_7^r=$ $\biggm\lvert\begin{array}{cccccc}
1 & 3    \\
1 & 2
\end{array}\biggm\rangle$}\vspace{0.05cm}\\ 
\hline
 \vspace{-0.05cm}$8$&\vspace{-0.05cm}$2.0$ &\vspace{-0.15cm}\resizebox{.2\textwidth}{!}{MI$_8^l=$ $\biggm\lvert\begin{array}{cccccc}
2 & 2    \\
2 & 2
\end{array}\biggm\rangle$ and MI$_8^r=$ $\biggm\lvert\begin{array}{cccccc}
1 & 3    \\
2 & 2
\end{array}\biggm\rangle$}\vspace{0.05cm}\\
\hline
\vspace{-0.05cm}$9$&\vspace{-0.05cm}$2.25$ &\vspace{-0.15cm}\resizebox{.1\textwidth}{!}{MI$_9=$ $\biggm\lvert\begin{array}{cccccc}
2 & 3    \\
2 & 2
\end{array}\biggm\rangle$}\vspace{0.05cm}\\ 
\hline
 \vspace{-0.05cm}$10$&\vspace{-0.05cm}$2.5$ &\vspace{-0.15cm}\resizebox{.2\textwidth}{!}{MI$_8^l=$ $\biggm\lvert\begin{array}{cccccc}
3 & 3    \\
2 & 2
\end{array}\biggm\rangle$ and MI$_8^r=$ $\biggm\lvert\begin{array}{cccccc}
2 & 4    \\
2 & 2
\end{array}\biggm\rangle$}\vspace{0.05cm}\\
\hline
\end{tabular}
\end{center}
\end{table}

 In Fig.~\ref{fig:rhomu_tbc} we show 
$\rho, \rho_s - \mu$ curves for $\lambda =10,~70,~140$ along the cuts through the gapped phases in Fig.~\ref{fig:cmftphasedia2}. 
As discussed in the previous section we see that for a particular $\lambda$, as the chemical potential increases the plateaus in $\rho$ appear 
and at the same time $\rho_s$ vanishes which correspond to the MI phases which are explained in Table-II.

It can be seen that the phase diagram for this case exhibits distinct features along with some similarities compared to the one obtained 
when bosons in layer-B was hardcore in nature(compare with Fig.~\ref{fig:cmftphasedia1}). 
Although the appearance of the MI phases for the odd (total)densities show similar behavior as the previous case, the even lobes 
exhibit different features with respect to their tip positions. It can be easily seen that the features in the part of the phase diagram from $n=6$ onwards 
matches well with the phase diagram of the previous case (when the layer-B was hardcore) except the first lobe at $n=1$ of Fig.~\ref{fig:cmftphasedia1}. This can be understood 
as follows. Let's consider the $\lambda=0$ case for simplicity. 
As the layer-B is occupied by TBCs now, for $n=6$ the density of the supercell is $1.5$ . This means, there are two 
extra particles on top of the $\rho=1$ lobe(MI$_4^l$). At $\rho=1$, each sites of both the layers are occupied by one 
particle  because of large onsite interactions $U_A$ and $U_B$. Therefore, any extra particle which gets 
added to layer-B containing the TBCs will behave like hardcore bosons on top of the uniform particle distribution. 
Hence, the effective system becomes equivalent to the one considered in the previous section. 
However, there exists different gapped phases at low density regimes i.e. upto $n=6$ lobes. For $n=1$($\rho=0.25$) we get the 
MI$_1$ phase which is similar to the one in Fig.\ref{fig:cmftphasedia1} which starts after finite value of $\lambda\sim 8$. 
As the number of particles increases, the second gapped phase MI$_2^l$ appears at $n=2$($\rho=0.5$) for $\lambda=0$
and this survives upto a critical value of $\lambda\sim32$ and after this the system becomes gapless. In this case, each layer is occupied by one 
boson. Further increase in $\lambda$ leads to the MI$_2^r$ phase, where the particles live in the deep sites. 
Similarly, the MI$_4$ phases appear for $n=4$($\rho=1$) in the beginning when $\lambda=0$ and the system is a proper Mott insulator 
at unit filling. Increase in $\lambda$ will melt the gap and the SF phase appears and further increase in $\lambda$ will
re-introduce the gap and the system gets into the  MI$_4^r$ phase. In this phase, there can be two possible particle distributions 
in the lattice where two particles populate the deep sites of layer-A while sites of layer-B are uniformly filled by one 
particle in each site and vice-versa. This gapped phase vanishes at $\lambda\sim 148$ and further increase in $\lambda$ may lead to another gapped 
phase where the deep sites of both the layers will be occupied by two particles(not in the range of $\lambda$ considered here). 
The physics of odd integer lobes can also be understood from the similar analogies discussed above.  

\begin{figure}[b]
  \centering
  \includegraphics[width=1\linewidth]{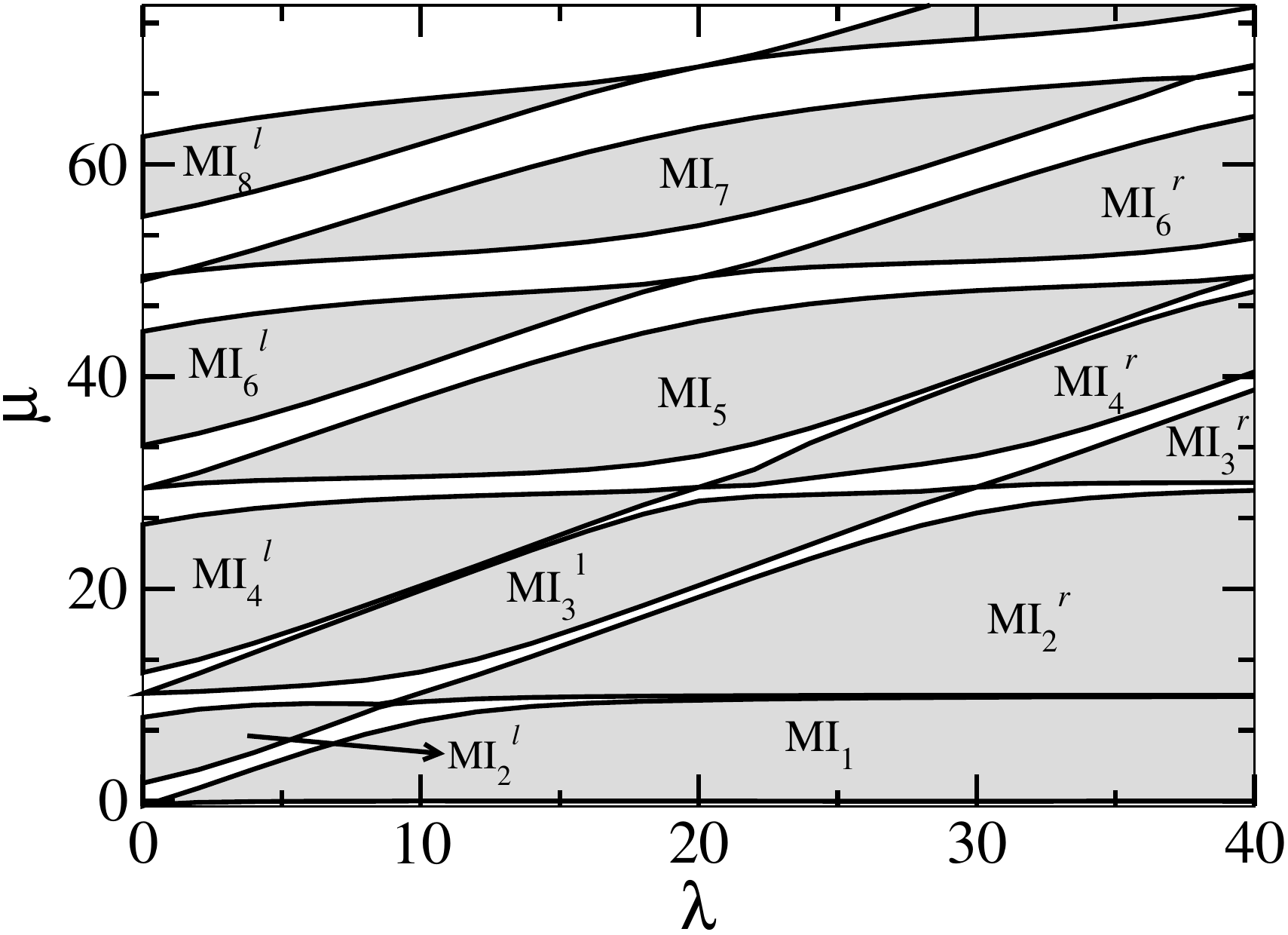}
  \caption{DMRG phase diagram when bosons in leg-A are softcore in nature and bosons in leg-B are HCBs.}
\label{fig:dmrg1}
\end{figure}

\begin{figure}[t]
  \centering
  \includegraphics[width=1\linewidth]{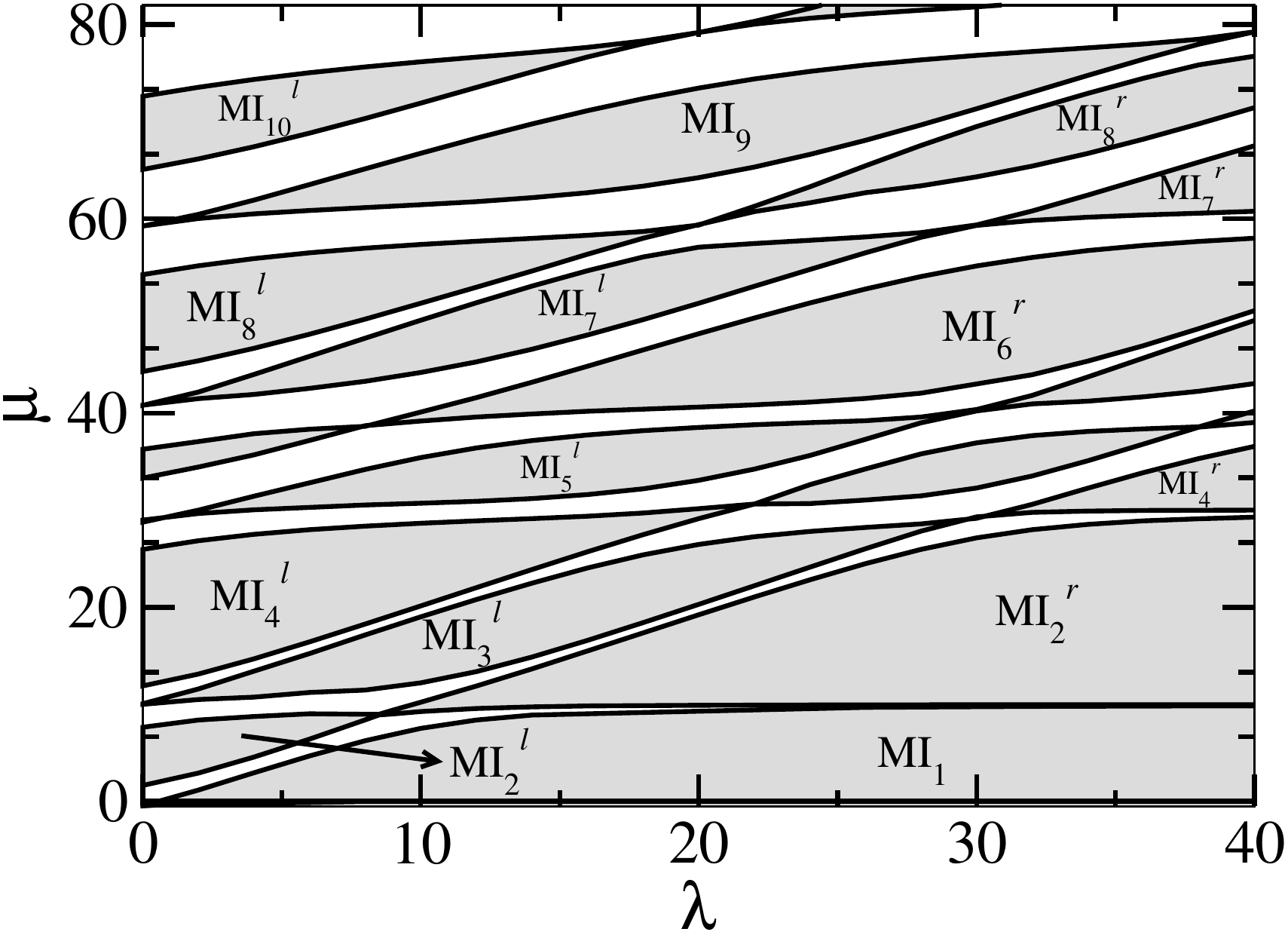}
  \caption{DMRG phase diagram when bosons in leg-A are softcore in nature and bosons in leg-B are TBCs.}
\label{fig:dmrg2}
\end{figure}

\begin{figure}[b]
  \centering
  \includegraphics[width=1\linewidth]{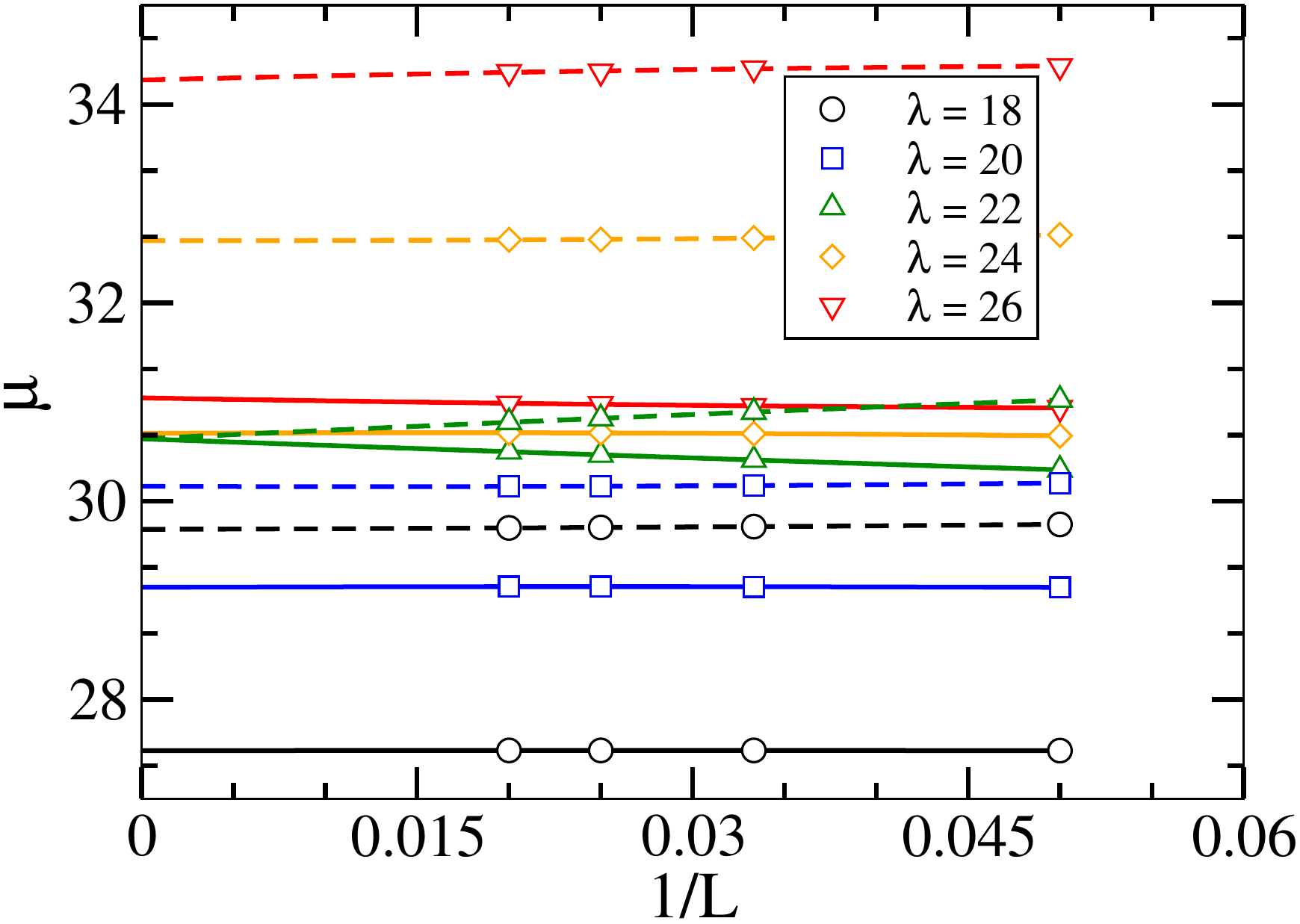}
  \caption{(Color online) Finite size scaling of chemical potentials for $\rho=1.5$ for different values of $\lambda$ corresponding to the phase diagram of Fig.~\ref{fig:dmrg1} when
  bosons in leg-A are softcore in nature and bosons in leg-B are HCBs. 
  The solid and dashed lines represent the fitted functions to $\mu^+$ and $\mu^-$ respectively.}
\label{fig:scalemu}
\end{figure}

\begin{figure}[t]
  \centering
  \includegraphics[width=1\linewidth]{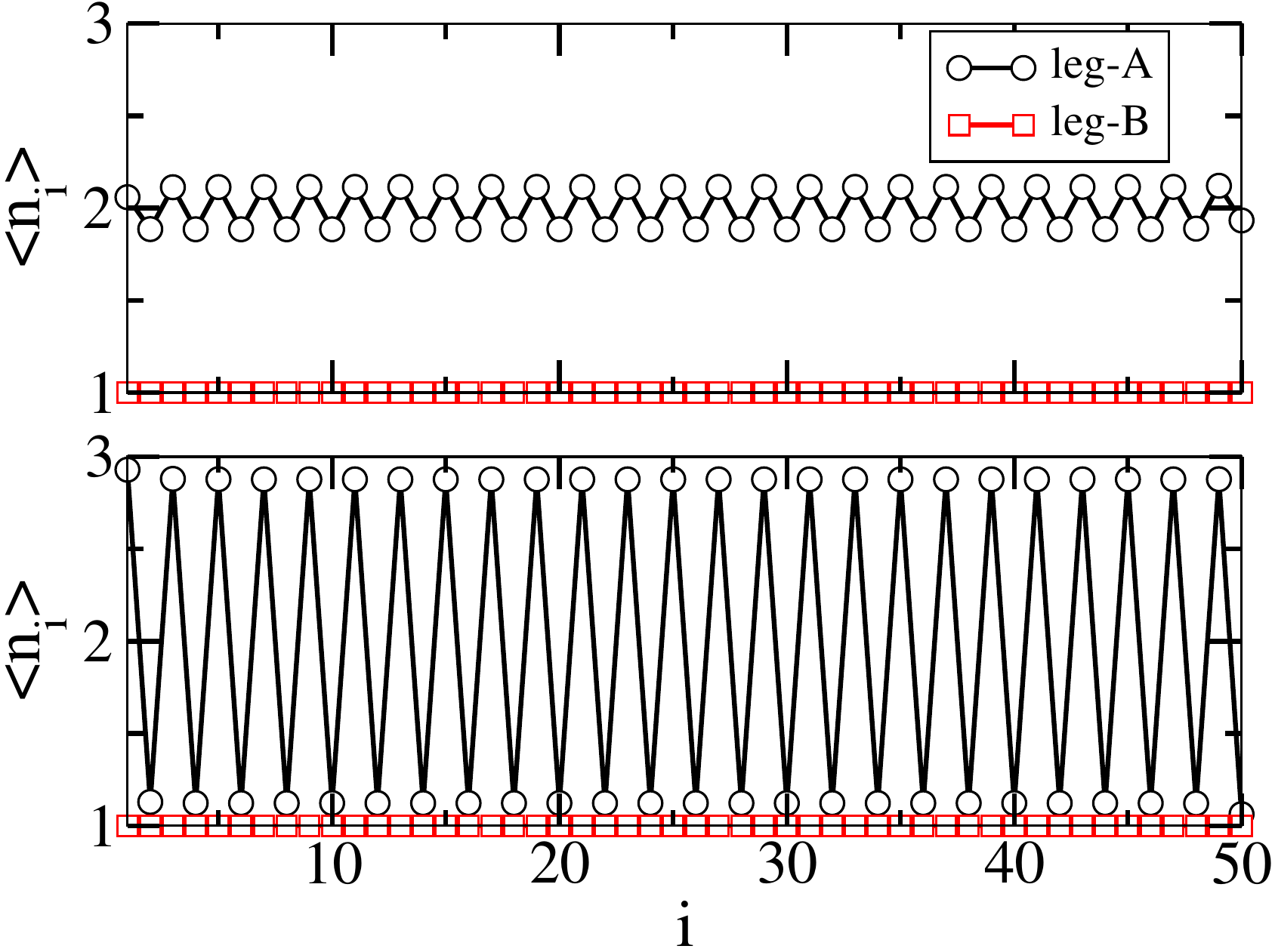}
  \caption{(Color online) Density distribution of the system when $\lambda=12$(top) and $\lambda=28$(bottom) for $\rho=1.5$ corresponding to the phase diagram of Fig.~\ref{fig:dmrg1}. 
  The black circles and red squares represent the density distribution of leg-A and leg-B respectively.}
\label{fig:dmrgdensityHC}
\end{figure}

\begin{figure}[b]
  \centering
  \includegraphics[width=1\linewidth]{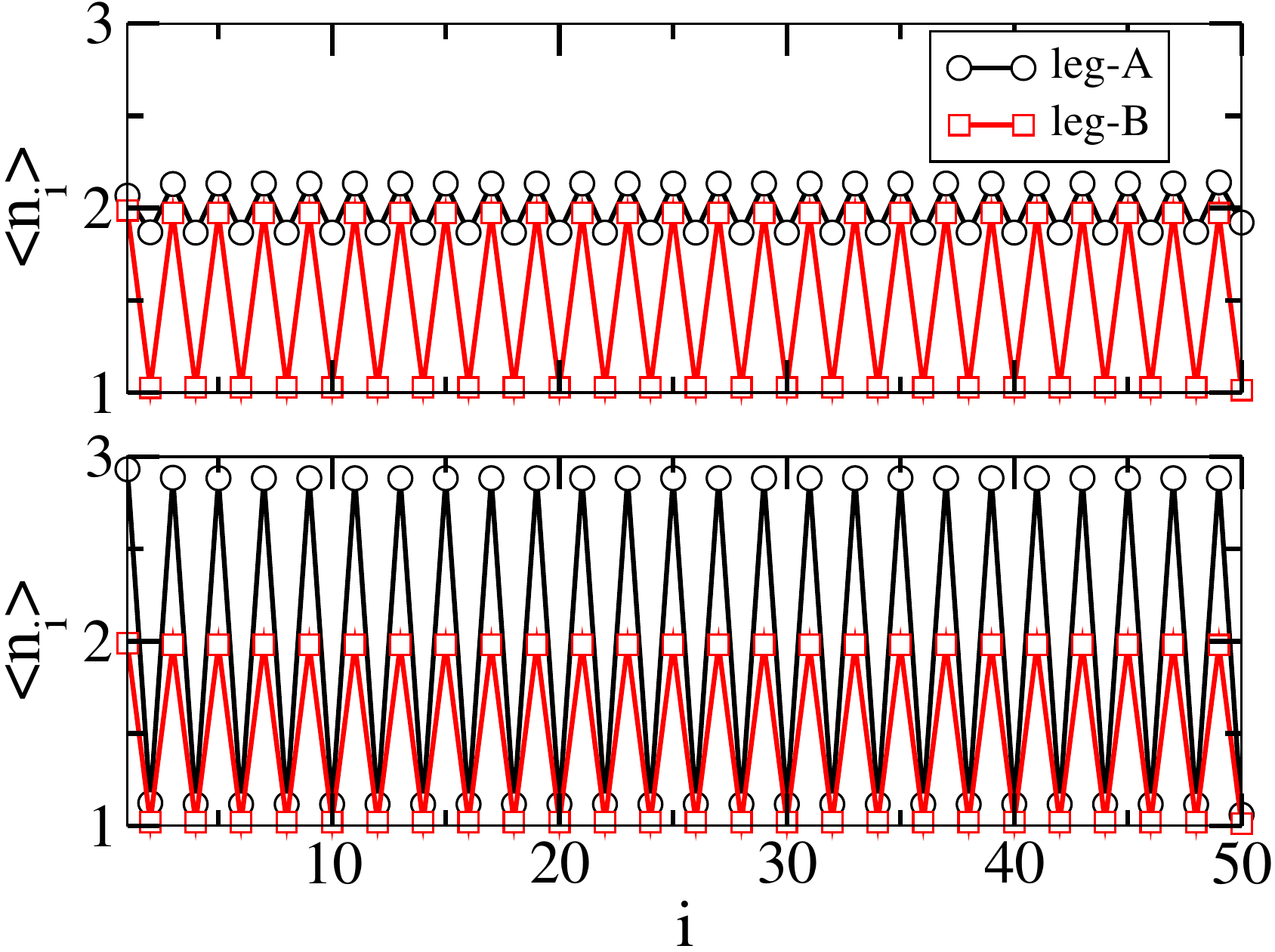}
  \caption{(Color online) Density distribution of the system when $\lambda=22$(top) and $\lambda=38$(bottom) for $\rho=1.75$ corresponding to the phase diagram of Fig.~\ref{fig:dmrg2}. 
  The black circles and red squares represent the density distribution of leg-A and leg-B respectively.}
\label{fig:dmrgdensityTBC}
\end{figure}

\subsection{Phase diagram in one dimension}
In this section we complement our CMFT results presented above 
by analyzing the situation in one dimension. The one dimensional analogue of bilayer geometry is a two-leg ladder where 
the layer-A(B) is replaced by leg-A(B) as shown in Fig.~\ref{fig:superlattice_with_period1}(c). The one dimensional ladder geometries are extremely important in 
the context of condensed matter systems as it resembles to several structures of compounds of interest. The ladder geometries has been 
discussed in great detail in terms of Hubbard model~\cite{danohue2001,FeiguinFisher2011,Simon2011,Okumura2011,Marco2016,Liu2019,giamarchi2003quantum,fabrizio,Nagaosa1995,schulz1996,
balents1996,Noack1996,white2002} and 
Bose-Hubbard model~\cite{Giamarchi2001,Tapan2008,santos_arturo,singh1,giamarchiladder,Cazalilla2006}. Analogous to the 2$d$ case we assume only inter-leg dipole-dipole interactions
by aligning the dipoles at magic angle with each other along the leg direction.
The physics of this system will be similar to the 
two dimensional case due to the construction of the bilayer lattice in our case as discussed earlier. 
We employ the DMRG method to solve the model(Eq.~\ref{eqn:ham}) in the canonical ensemble  
to compute the ground state energy and wave function. To separate the gapped and gapless regions we 
calculate the single particle gap which is defined as 
\begin{equation}
 G_L=\mu^+-\mu^-,
\end{equation}
with $\mu^+=E_L(N+1)-E_L(N)$ and $\mu^-=E_L(N)-E_L(N-1)$ are the chemical potentials and $E_L(N)$ is the 
the ground state energy of a system of length $L$ and $N=N_A+N_B$ is the total number of particles in the system . 
We obtain the ground state phase diagram for both the cases with bosons of leg-B separately being HCBs and TBCs while bosons 
in leg-A are softcore in nature which are shown 
in Fig.~\ref{fig:dmrg1}($U_{A}=20$ and $U_{AB}=10$) and Fig.~\ref{fig:dmrg2}($U_{A,B}=20$ and $U_{AB}=10$) respectively. 
One can 
easily see that the phase diagrams of Fig.~\ref{fig:dmrg1} and Fig.~\ref{fig:dmrg2} qualitatively match fairly well with the ones obtained using 
the CMFT method i.e. Fig. \ref{fig:cmftphasedia1} 
and Fig. \ref{fig:cmftphasedia2} respectively. 
When bosons in leg-B are HCBs, we observe the MI tip positions first increase 
and then decrease as shown in Fig.~\ref{fig:dmrg1} while there is an alternating 
increase and decrease of the tip positions when the bosons in leg-B are TBCs as plotted in Fig.~\ref{fig:dmrg2}. 
The boundaries of the MI lobes are computed by extrapolating the $\mu$ values across the MI plateaus to thermodynamic limit by quadratic 
fitting. In Fig. ~\ref{fig:scalemu}, we show the finite size extrapolation of $\mu^+$(dashed) and $\mu^-$(solid) 
for $\lambda=18,~20,~22,~24$ and $26$. This clearly shows that $G_{L\longrightarrow\infty}$ remains finite in the gapped phase and vanishes in the gapless region which
clearly distinguish between the gapped and gapless phases in the phase diagrams.

Further, to understand the particle distribution in real space we compute the expectation value of the number operator as 
\begin{equation}
 \langle n_i\rangle =\langle \Psi_0| n_i| \Psi_0 \rangle
\end{equation}
where $|\Psi_0\rangle$ is the ground state wave function of the system. As an example, in Fig.~\ref{fig:dmrgdensityHC} we plot $\langle n_i\rangle$ w.r.t. site index $i$ 
for $\lambda=12$(top) and $\lambda=28$(bottom) corresponding to the MI$_6^l$ and MI$_6^r$ phases respectively of the phase diagram shown in Fig.~\ref{fig:dmrg1}. 
It can be clearly seen that for $\lambda=12$ the leg-B is occupied by one hardcore boson in each site whereas in leg-A, each site is occupied by two atoms.
However, for $\lambda=28$, the leg-B is unaffected and leg-A shows $|.~.~3~1~3~1~.~.\rangle$ type of distribution corresponding to the MI$_6^r$ phase.
A similar $\langle n_i\rangle$ $vs$ $i$ plot is shown in Fig.~\ref{fig:dmrgdensityTBC}
for the MI$_7$ phase of Fig.~\ref{fig:dmrg2}. Here we consider $\lambda = 22$(top) and $38$(bottom) which fall in two regions of the MI$_7^l$ and MI$_7^r$ phases respectively. 
For $\lambda=22$, all the sites of leg-A are occupied by two particles each whereas leg-B exhibits a finite density oscillation corresponding to $|.~.~2~1~2~1~.~.\rangle$ type of distribution.
However, for large $\lambda=38$,
the density distribution of leg-A becomes $|.~.~3~1~3~1~.~.\rangle$ while the leg-B remains unaffected.
It is to be noted that for the parameters 
considered here, the gapless regions between two gapped phases are very small compared to ones obtained using the 
CMFT method.

\section{conclusion}

We analyse the ground state properties of a system of interacting bosons in bilayer superlattice with inter-layer repulsion which 
can be introduced by the dipole-dipole interactions. 
Considering the bosons in one layer as softcore in nature and separately allowing two and three-body hardcore constraints 
in the other layer we obtain the ground state phase diagram using the CMFT approach. The phase diagrams exhibit 
various gapped MI phases at integer and half-integer densities. Due to the competition between the superlattice potential, intra- and inter-layer interactions 
and the constraints on the bosons of layer-B leads to interesting features in the phase diagram. 
It is shown that within the range of $\lambda$ considered, we obtain two types MI lobes(MI$^l_n$ and MI$^r_n$) 
separated by the SF region for a particular total number of particles in the supercell equal to $n$.  
Interestingly, when the hardcore constraint is applied in one layer, 
the tips of the MI$^l_n$ lobes first shift towards higher values of $\lambda$ and 
then gradually recede to the lower values of $\lambda$. At the same 
time the tips of the MI$^r_n$ lobes shift towards the higher $\lambda$ values with increase in density of the system. 
The situation is completely different when the three-body constraint is considered in one layer. The 
tips of MI lobes first oscillate and then after a critical density follows the trend similar to the one for the hardcore 
constraint. We further complement our findings by repeating the calculations in an one dimensional non-locally coupled 
ladder superlattice using the DMRG method and show that the quantum phase diagrams qualitatively agree with the CMFT method. 
The physics obtained in this work deals with the system of bosons in two-layer systems with different types of 
onsite interactions in a superlattice. The results provide detailed analysis of the effect of constrained bosons on the 
overall phase diagram of the bilayer system which is also equivalent to a two-component atomic systems. With the experimental 
progress in controlling local and dipole-dipole interactions in recent years, these findings can be experimentally observed 
with the existing quantum gas setups. 

\section{Acknowledgement}
We thank B. P. Das for useful discussion. 
The computational works were carried out using the
Param-Ishan HPC facility at the Indian Institute of Technology - Guwahati, India. T.M. acknowledges 
DST-SERB for the early career grant through Project No.
ECR/2017/001069 and the hospitality received from ICTS-TIFR, Bangalore where part of the manuscript was written. 
\bibliography{paper3_ref}

\begin{thebibliography}{82}%
\makeatletter
\providecommand \@ifxundefined [1]{%
 \@ifx{#1\undefined}
}%
\providecommand \@ifnum [1]{%
 \ifnum #1\expandafter \@firstoftwo
 \else \expandafter \@secondoftwo
 \fi
}%
\providecommand \@ifx [1]{%
 \ifx #1\expandafter \@firstoftwo
 \else \expandafter \@secondoftwo
 \fi
}%
\providecommand \natexlab [1]{#1}%
\providecommand \enquote  [1]{``#1''}%
\providecommand \bibnamefont  [1]{#1}%
\providecommand \bibfnamefont [1]{#1}%
\providecommand \citenamefont [1]{#1}%
\providecommand \href@noop [0]{\@secondoftwo}%
\providecommand \href [0]{\begingroup \@sanitize@url \@href}%
\providecommand \@href[1]{\@@startlink{#1}\@@href}%
\providecommand \@@href[1]{\endgroup#1\@@endlink}%
\providecommand \@sanitize@url [0]{\catcode `\\12\catcode `\$12\catcode
  `\&12\catcode `\#12\catcode `\^12\catcode `\_12\catcode `\%12\relax}%
\providecommand \@@startlink[1]{}%
\providecommand \@@endlink[0]{}%
\providecommand \url  [0]{\begingroup\@sanitize@url \@url }%
\providecommand \@url [1]{\endgroup\@href {#1}{\urlprefix }}%
\providecommand \urlprefix  [0]{URL }%
\providecommand \Eprint [0]{\href }%
\providecommand \doibase [0]{http://dx.doi.org/}%
\providecommand \selectlanguage [0]{\@gobble}%
\providecommand \bibinfo  [0]{\@secondoftwo}%
\providecommand \bibfield  [0]{\@secondoftwo}%
\providecommand \translation [1]{[#1]}%
\providecommand \BibitemOpen [0]{}%
\providecommand \bibitemStop [0]{}%
\providecommand \bibitemNoStop [0]{.\EOS\space}%
\providecommand \EOS [0]{\spacefactor3000\relax}%
\providecommand \BibitemShut  [1]{\csname bibitem#1\endcsname}%
\let\auto@bib@innerbib\@empty
\bibitem [{\citenamefont {Greiner}\ \emph {et~al.}(2002)\citenamefont
  {Greiner}, \citenamefont {Mandel}, \citenamefont {Esslinger}, \citenamefont
  {H{\"a}nsch},\ and\ \citenamefont {Bloch}}]{bloch}%
  \BibitemOpen
  \bibfield  {author} {\bibinfo {author} {\bibfnamefont {M.}~\bibnamefont
  {Greiner}}, \bibinfo {author} {\bibfnamefont {O.}~\bibnamefont {Mandel}},
  \bibinfo {author} {\bibfnamefont {T.}~\bibnamefont {Esslinger}}, \bibinfo
  {author} {\bibfnamefont {T.~W.}\ \bibnamefont {H{\"a}nsch}}, \ and\ \bibinfo
  {author} {\bibfnamefont {I.}~\bibnamefont {Bloch}},\ }\href {\doibase
  https://doi.org/10.1038/415039a} {\bibfield  {journal} {\bibinfo  {journal}
  {Nature}\ }\textbf {\bibinfo {volume} {415}},\ \bibinfo {pages} {39 EP }
  (\bibinfo {year} {2002})},\ \bibinfo {note} {article}\BibitemShut {NoStop}%
\bibitem [{\citenamefont {Fisher}\ \emph {et~al.}(1989)\citenamefont {Fisher},
  \citenamefont {Weichman}, \citenamefont {Grinstein},\ and\ \citenamefont
  {Fisher}}]{FisherPRB1989}%
  \BibitemOpen
  \bibfield  {author} {\bibinfo {author} {\bibfnamefont {M.~P.~A.}\
  \bibnamefont {Fisher}}, \bibinfo {author} {\bibfnamefont {P.~B.}\
  \bibnamefont {Weichman}}, \bibinfo {author} {\bibfnamefont {G.}~\bibnamefont
  {Grinstein}}, \ and\ \bibinfo {author} {\bibfnamefont {D.~S.}\ \bibnamefont
  {Fisher}},\ }\href {\doibase 10.1103/PhysRevB.40.546} {\bibfield  {journal}
  {\bibinfo  {journal} {Phys. Rev. B}\ }\textbf {\bibinfo {volume} {40}},\
  \bibinfo {pages} {546} (\bibinfo {year} {1989})}\BibitemShut {NoStop}%
\bibitem [{\citenamefont {Jaksch}\ \emph {et~al.}(1998)\citenamefont {Jaksch},
  \citenamefont {Bruder}, \citenamefont {Cirac}, \citenamefont {Gardiner},\
  and\ \citenamefont {Zoller}}]{jaksch}%
  \BibitemOpen
  \bibfield  {author} {\bibinfo {author} {\bibfnamefont {D.}~\bibnamefont
  {Jaksch}}, \bibinfo {author} {\bibfnamefont {C.}~\bibnamefont {Bruder}},
  \bibinfo {author} {\bibfnamefont {J.~I.}\ \bibnamefont {Cirac}}, \bibinfo
  {author} {\bibfnamefont {C.~W.}\ \bibnamefont {Gardiner}}, \ and\ \bibinfo
  {author} {\bibfnamefont {P.}~\bibnamefont {Zoller}},\ }\href {\doibase
  10.1103/PhysRevLett.81.3108} {\bibfield  {journal} {\bibinfo  {journal}
  {Phys. Rev. Lett.}\ }\textbf {\bibinfo {volume} {81}},\ \bibinfo {pages}
  {3108} (\bibinfo {year} {1998})}\BibitemShut {NoStop}%
\bibitem [{\citenamefont {{S. Will}}\ \emph {et~al.}(2010)\citenamefont {{S.
  Will}}, \citenamefont {{T. Best}}, \citenamefont {{U. Schneider}},
  \citenamefont {{L. Hackerm{\"u}ller}}, \citenamefont {{D.S. L{\"u}hmann}},\
  and\ \citenamefont {{I. Bloch}}}]{wills}%
  \BibitemOpen
  \bibfield  {author} {\bibinfo {author} {\bibnamefont {{S. Will}}}, \bibinfo
  {author} {\bibnamefont {{T. Best}}}, \bibinfo {author} {\bibnamefont {{U.
  Schneider}}}, \bibinfo {author} {\bibnamefont {{L. Hackerm{\"u}ller}}},
  \bibinfo {author} {\bibnamefont {{D.S. L{\"u}hmann}}}, \ and\ \bibinfo
  {author} {\bibnamefont {{I. Bloch}}},\ }\href {\doibase
  https://doi.org/10.1038/nature09036 10.1038/nature09036} {\bibfield
  {journal} {\bibinfo  {journal} {Nature}\ }\textbf {\bibinfo {volume} {465}},\
  \bibinfo {pages} {197} (\bibinfo {year} {2010})}\BibitemShut {NoStop}%
\bibitem [{\citenamefont {Mark}\ \emph {et~al.}(2011)\citenamefont {Mark},
  \citenamefont {Haller}, \citenamefont {Lauber}, \citenamefont {Danzl},
  \citenamefont {Daley},\ and\ \citenamefont {N\"agerl}}]{mark2011}%
  \BibitemOpen
  \bibfield  {author} {\bibinfo {author} {\bibfnamefont {M.~J.}\ \bibnamefont
  {Mark}}, \bibinfo {author} {\bibfnamefont {E.}~\bibnamefont {Haller}},
  \bibinfo {author} {\bibfnamefont {K.}~\bibnamefont {Lauber}}, \bibinfo
  {author} {\bibfnamefont {J.~G.}\ \bibnamefont {Danzl}}, \bibinfo {author}
  {\bibfnamefont {A.~J.}\ \bibnamefont {Daley}}, \ and\ \bibinfo {author}
  {\bibfnamefont {H.-C.}\ \bibnamefont {N\"agerl}},\ }\href {\doibase
  10.1103/PhysRevLett.107.175301} {\bibfield  {journal} {\bibinfo  {journal}
  {Phys. Rev. Lett.}\ }\textbf {\bibinfo {volume} {107}},\ \bibinfo {pages}
  {175301} (\bibinfo {year} {2011})}\BibitemShut {NoStop}%
\bibitem [{\citenamefont {Daley}\ and\ \citenamefont {Simon}(2014)}]{daley}%
  \BibitemOpen
  \bibfield  {author} {\bibinfo {author} {\bibfnamefont {A.~J.}\ \bibnamefont
  {Daley}}\ and\ \bibinfo {author} {\bibfnamefont {J.}~\bibnamefont {Simon}},\
  }\href {\doibase 10.1103/PhysRevA.89.053619} {\bibfield  {journal} {\bibinfo
  {journal} {Phys. Rev. A}\ }\textbf {\bibinfo {volume} {89}},\ \bibinfo
  {pages} {053619} (\bibinfo {year} {2014})}\BibitemShut {NoStop}%
\bibitem [{\citenamefont {Johnson}\ \emph {et~al.}(2009)\citenamefont
  {Johnson}, \citenamefont {Tiesinga}, \citenamefont {Porto},\ and\
  \citenamefont {Williams}}]{tiesinga}%
  \BibitemOpen
  \bibfield  {author} {\bibinfo {author} {\bibfnamefont {P.~R.}\ \bibnamefont
  {Johnson}}, \bibinfo {author} {\bibfnamefont {E.}~\bibnamefont {Tiesinga}},
  \bibinfo {author} {\bibfnamefont {J.~V.}\ \bibnamefont {Porto}}, \ and\
  \bibinfo {author} {\bibfnamefont {C.~J.}\ \bibnamefont {Williams}},\ }\href
  {https://iopscience.iop.org/article/10.1088/1367-2630/11/9/093022/meta}
  {\bibfield  {journal} {\bibinfo  {journal} {New Journal of Physics}\ }\textbf
  {\bibinfo {volume} {11}},\ \bibinfo {pages} {093022} (\bibinfo {year}
  {2009})}\BibitemShut {NoStop}%
\bibitem [{\citenamefont {Petrov}(2014{\natexlab{a}})}]{petrov1}%
  \BibitemOpen
  \bibfield  {author} {\bibinfo {author} {\bibfnamefont {D.~S.}\ \bibnamefont
  {Petrov}},\ }\href {\doibase 10.1103/PhysRevLett.112.103201} {\bibfield
  {journal} {\bibinfo  {journal} {Phys. Rev. Lett.}\ }\textbf {\bibinfo
  {volume} {112}},\ \bibinfo {pages} {103201} (\bibinfo {year}
  {2014}{\natexlab{a}})}\BibitemShut {NoStop}%
\bibitem [{\citenamefont {Petrov}(2014{\natexlab{b}})}]{petrov2}%
  \BibitemOpen
  \bibfield  {author} {\bibinfo {author} {\bibfnamefont {D.~S.}\ \bibnamefont
  {Petrov}},\ }\href {\doibase 10.1103/PhysRevA.90.021601} {\bibfield
  {journal} {\bibinfo  {journal} {Phys. Rev. A}\ }\textbf {\bibinfo {volume}
  {90}},\ \bibinfo {pages} {021601} (\bibinfo {year}
  {2014}{\natexlab{b}})}\BibitemShut {NoStop}%
\bibitem [{\citenamefont {Singh}\ \emph
  {et~al.}(2012{\natexlab{a}})\citenamefont {Singh}, \citenamefont {Dhar},
  \citenamefont {Mishra}, \citenamefont {Pai},\ and\ \citenamefont
  {Das}}]{manpreetsuplat}%
  \BibitemOpen
  \bibfield  {author} {\bibinfo {author} {\bibfnamefont {M.}~\bibnamefont
  {Singh}}, \bibinfo {author} {\bibfnamefont {A.}~\bibnamefont {Dhar}},
  \bibinfo {author} {\bibfnamefont {T.}~\bibnamefont {Mishra}}, \bibinfo
  {author} {\bibfnamefont {R.~V.}\ \bibnamefont {Pai}}, \ and\ \bibinfo
  {author} {\bibfnamefont {B.~P.}\ \bibnamefont {Das}},\ }\href {\doibase
  10.1103/PhysRevA.85.051604} {\bibfield  {journal} {\bibinfo  {journal} {Phys.
  Rev. A}\ }\textbf {\bibinfo {volume} {85}},\ \bibinfo {pages} {051604}
  (\bibinfo {year} {2012}{\natexlab{a}})}\BibitemShut {NoStop}%
\bibitem [{\citenamefont {Sowi\ifmmode~\acute{n}\else \'{n}\fi{}ski}\ \emph
  {et~al.}(2015)\citenamefont {Sowi\ifmmode~\acute{n}\else \'{n}\fi{}ski},
  \citenamefont {Chhajlany}, \citenamefont {Dutta}, \citenamefont
  {Tagliacozzo},\ and\ \citenamefont {Lewenstein}}]{sowinski1}%
  \BibitemOpen
  \bibfield  {author} {\bibinfo {author} {\bibfnamefont {T.}~\bibnamefont
  {Sowi\ifmmode~\acute{n}\else \'{n}\fi{}ski}}, \bibinfo {author}
  {\bibfnamefont {R.~W.}\ \bibnamefont {Chhajlany}}, \bibinfo {author}
  {\bibfnamefont {O.}~\bibnamefont {Dutta}}, \bibinfo {author} {\bibfnamefont
  {L.}~\bibnamefont {Tagliacozzo}}, \ and\ \bibinfo {author} {\bibfnamefont
  {M.}~\bibnamefont {Lewenstein}},\ }\href {\doibase
  10.1103/PhysRevA.92.043615} {\bibfield  {journal} {\bibinfo  {journal} {Phys.
  Rev. A}\ }\textbf {\bibinfo {volume} {92}},\ \bibinfo {pages} {043615}
  (\bibinfo {year} {2015})}\BibitemShut {NoStop}%
\bibitem [{\citenamefont {Daley}\ \emph {et~al.}(2009)\citenamefont {Daley},
  \citenamefont {Taylor}, \citenamefont {Diehl}, \citenamefont {Baranov},\ and\
  \citenamefont {Zoller}}]{baranovprl}%
  \BibitemOpen
  \bibfield  {author} {\bibinfo {author} {\bibfnamefont {A.~J.}\ \bibnamefont
  {Daley}}, \bibinfo {author} {\bibfnamefont {J.~M.}\ \bibnamefont {Taylor}},
  \bibinfo {author} {\bibfnamefont {S.}~\bibnamefont {Diehl}}, \bibinfo
  {author} {\bibfnamefont {M.}~\bibnamefont {Baranov}}, \ and\ \bibinfo
  {author} {\bibfnamefont {P.}~\bibnamefont {Zoller}},\ }\href {\doibase
  10.1103/PhysRevLett.102.040402} {\bibfield  {journal} {\bibinfo  {journal}
  {Phys. Rev. Lett.}\ }\textbf {\bibinfo {volume} {102}},\ \bibinfo {pages}
  {040402} (\bibinfo {year} {2009})}\BibitemShut {NoStop}%
\bibitem [{\citenamefont {Paredes}\ \emph {et~al.}(2004)\citenamefont
  {Paredes}, \citenamefont {Widera}, \citenamefont {Murg}, \citenamefont
  {Mandel}, \citenamefont {F{\"o}lling}, \citenamefont {Cirac}, \citenamefont
  {Shlyapnikov}, \citenamefont {H{\"a}nsch},\ and\ \citenamefont
  {Bloch}}]{paredes1}%
  \BibitemOpen
  \bibfield  {author} {\bibinfo {author} {\bibfnamefont {B.}~\bibnamefont
  {Paredes}}, \bibinfo {author} {\bibfnamefont {A.}~\bibnamefont {Widera}},
  \bibinfo {author} {\bibfnamefont {V.}~\bibnamefont {Murg}}, \bibinfo {author}
  {\bibfnamefont {O.}~\bibnamefont {Mandel}}, \bibinfo {author} {\bibfnamefont
  {S.}~\bibnamefont {F{\"o}lling}}, \bibinfo {author} {\bibfnamefont
  {I.}~\bibnamefont {Cirac}}, \bibinfo {author} {\bibfnamefont {G.~V.}\
  \bibnamefont {Shlyapnikov}}, \bibinfo {author} {\bibfnamefont {T.~W.}\
  \bibnamefont {H{\"a}nsch}}, \ and\ \bibinfo {author} {\bibfnamefont
  {I.}~\bibnamefont {Bloch}},\ }\href
  {https://www.nature.com/articles/nature02530} {\bibfield  {journal} {\bibinfo
   {journal} {Nature}\ }\textbf {\bibinfo {volume} {429}},\ \bibinfo {pages}
  {277} (\bibinfo {year} {2004})}\BibitemShut {NoStop}%
\bibitem [{\citenamefont {Baranov}\ \emph {et~al.}(2012)\citenamefont
  {Baranov}, \citenamefont {Dalmonte}, \citenamefont {Pupillo},\ and\
  \citenamefont {Zoller}}]{baranovrev}%
  \BibitemOpen
  \bibfield  {author} {\bibinfo {author} {\bibfnamefont {M.~A.}\ \bibnamefont
  {Baranov}}, \bibinfo {author} {\bibfnamefont {M.}~\bibnamefont {Dalmonte}},
  \bibinfo {author} {\bibfnamefont {G.}~\bibnamefont {Pupillo}}, \ and\
  \bibinfo {author} {\bibfnamefont {P.}~\bibnamefont {Zoller}},\ }\href
  {\doibase 10.1021/cr2003568} {\bibfield  {journal} {\bibinfo  {journal}
  {Chemical Reviews}\ }\textbf {\bibinfo {volume} {112}},\ \bibinfo {pages}
  {5012} (\bibinfo {year} {2012})}\BibitemShut {NoStop}%
\bibitem [{\citenamefont {Baier}\ \emph {et~al.}(2016)\citenamefont {Baier},
  \citenamefont {Mark}, \citenamefont {Petter}, \citenamefont {Aikawa},
  \citenamefont {Chomaz}, \citenamefont {Cai}, \citenamefont {Baranov},
  \citenamefont {Zoller},\ and\ \citenamefont {Ferlaino}}]{Ferlaino1}%
  \BibitemOpen
  \bibfield  {author} {\bibinfo {author} {\bibfnamefont {S.}~\bibnamefont
  {Baier}}, \bibinfo {author} {\bibfnamefont {M.~J.}\ \bibnamefont {Mark}},
  \bibinfo {author} {\bibfnamefont {D.}~\bibnamefont {Petter}}, \bibinfo
  {author} {\bibfnamefont {K.}~\bibnamefont {Aikawa}}, \bibinfo {author}
  {\bibfnamefont {L.}~\bibnamefont {Chomaz}}, \bibinfo {author} {\bibfnamefont
  {Z.}~\bibnamefont {Cai}}, \bibinfo {author} {\bibfnamefont {M.}~\bibnamefont
  {Baranov}}, \bibinfo {author} {\bibfnamefont {P.}~\bibnamefont {Zoller}}, \
  and\ \bibinfo {author} {\bibfnamefont {F.}~\bibnamefont {Ferlaino}},\ }\href
  {\doibase 10.1126/science.aac9812} {\bibfield  {journal} {\bibinfo  {journal}
  {Science}\ }\textbf {\bibinfo {volume} {352}},\ \bibinfo {pages} {201}
  (\bibinfo {year} {2016})}\BibitemShut {NoStop}%
\bibitem [{\citenamefont {Chomaz}\ \emph {et~al.}(2019)\citenamefont {Chomaz},
  \citenamefont {Petter}, \citenamefont {Ilzh\"ofer}, \citenamefont {Natale},
  \citenamefont {Trautmann}, \citenamefont {Politi}, \citenamefont
  {Durastante}, \citenamefont {van Bijnen}, \citenamefont {Patscheider},
  \citenamefont {Sohmen}, \citenamefont {Mark},\ and\ \citenamefont
  {Ferlaino}}]{Ferlaino2}%
  \BibitemOpen
  \bibfield  {author} {\bibinfo {author} {\bibfnamefont {L.}~\bibnamefont
  {Chomaz}}, \bibinfo {author} {\bibfnamefont {D.}~\bibnamefont {Petter}},
  \bibinfo {author} {\bibfnamefont {P.}~\bibnamefont {Ilzh\"ofer}}, \bibinfo
  {author} {\bibfnamefont {G.}~\bibnamefont {Natale}}, \bibinfo {author}
  {\bibfnamefont {A.}~\bibnamefont {Trautmann}}, \bibinfo {author}
  {\bibfnamefont {C.}~\bibnamefont {Politi}}, \bibinfo {author} {\bibfnamefont
  {G.}~\bibnamefont {Durastante}}, \bibinfo {author} {\bibfnamefont {R.~M.~W.}\
  \bibnamefont {van Bijnen}}, \bibinfo {author} {\bibfnamefont
  {A.}~\bibnamefont {Patscheider}}, \bibinfo {author} {\bibfnamefont
  {M.}~\bibnamefont {Sohmen}}, \bibinfo {author} {\bibfnamefont {M.~J.}\
  \bibnamefont {Mark}}, \ and\ \bibinfo {author} {\bibfnamefont
  {F.}~\bibnamefont {Ferlaino}},\ }\href {\doibase 10.1103/PhysRevX.9.021012}
  {\bibfield  {journal} {\bibinfo  {journal} {Phys. Rev. X}\ }\textbf {\bibinfo
  {volume} {9}},\ \bibinfo {pages} {021012} (\bibinfo {year}
  {2019})}\BibitemShut {NoStop}%
\bibitem [{\citenamefont {Tanzi}\ \emph {et~al.}(2019)\citenamefont {Tanzi},
  \citenamefont {Lucioni}, \citenamefont {Fam\`a}, \citenamefont {Catani},
  \citenamefont {Fioretti}, \citenamefont {Gabbanini}, \citenamefont {Bisset},
  \citenamefont {Santos},\ and\ \citenamefont {Modugno}}]{tanzi2019}%
  \BibitemOpen
  \bibfield  {author} {\bibinfo {author} {\bibfnamefont {L.}~\bibnamefont
  {Tanzi}}, \bibinfo {author} {\bibfnamefont {E.}~\bibnamefont {Lucioni}},
  \bibinfo {author} {\bibfnamefont {F.}~\bibnamefont {Fam\`a}}, \bibinfo
  {author} {\bibfnamefont {J.}~\bibnamefont {Catani}}, \bibinfo {author}
  {\bibfnamefont {A.}~\bibnamefont {Fioretti}}, \bibinfo {author}
  {\bibfnamefont {C.}~\bibnamefont {Gabbanini}}, \bibinfo {author}
  {\bibfnamefont {R.~N.}\ \bibnamefont {Bisset}}, \bibinfo {author}
  {\bibfnamefont {L.}~\bibnamefont {Santos}}, \ and\ \bibinfo {author}
  {\bibfnamefont {G.}~\bibnamefont {Modugno}},\ }\href {\doibase
  10.1103/PhysRevLett.122.130405} {\bibfield  {journal} {\bibinfo  {journal}
  {Phys. Rev. Lett.}\ }\textbf {\bibinfo {volume} {122}},\ \bibinfo {pages}
  {130405} (\bibinfo {year} {2019})}\BibitemShut {NoStop}%
\bibitem [{\citenamefont {B\"ottcher}\ \emph {et~al.}(2019)\citenamefont
  {B\"ottcher}, \citenamefont {Schmidt}, \citenamefont {Wenzel}, \citenamefont
  {Hertkorn}, \citenamefont {Guo}, \citenamefont {Langen},\ and\ \citenamefont
  {Pfau}}]{bottcher2019}%
  \BibitemOpen
  \bibfield  {author} {\bibinfo {author} {\bibfnamefont {F.}~\bibnamefont
  {B\"ottcher}}, \bibinfo {author} {\bibfnamefont {J.-N.}\ \bibnamefont
  {Schmidt}}, \bibinfo {author} {\bibfnamefont {M.}~\bibnamefont {Wenzel}},
  \bibinfo {author} {\bibfnamefont {J.}~\bibnamefont {Hertkorn}}, \bibinfo
  {author} {\bibfnamefont {M.}~\bibnamefont {Guo}}, \bibinfo {author}
  {\bibfnamefont {T.}~\bibnamefont {Langen}}, \ and\ \bibinfo {author}
  {\bibfnamefont {T.}~\bibnamefont {Pfau}},\ }\href {\doibase
  10.1103/PhysRevX.9.011051} {\bibfield  {journal} {\bibinfo  {journal} {Phys.
  Rev. X}\ }\textbf {\bibinfo {volume} {9}},\ \bibinfo {pages} {011051}
  (\bibinfo {year} {2019})}\BibitemShut {NoStop}%
\bibitem [{\citenamefont {Trefzger}\ \emph {et~al.}(2009)\citenamefont
  {Trefzger}, \citenamefont {Menotti},\ and\ \citenamefont
  {Lewenstein}}]{lewenstein}%
  \BibitemOpen
  \bibfield  {author} {\bibinfo {author} {\bibfnamefont {C.}~\bibnamefont
  {Trefzger}}, \bibinfo {author} {\bibfnamefont {C.}~\bibnamefont {Menotti}}, \
  and\ \bibinfo {author} {\bibfnamefont {M.}~\bibnamefont {Lewenstein}},\
  }\href {\doibase 10.1103/PhysRevLett.103.035304} {\bibfield  {journal}
  {\bibinfo  {journal} {Phys. Rev. Lett.}\ }\textbf {\bibinfo {volume} {103}},\
  \bibinfo {pages} {035304} (\bibinfo {year} {2009})}\BibitemShut {NoStop}%
\bibitem [{\citenamefont {Safavi-Naini}\ \emph {et~al.}(2013)\citenamefont
  {Safavi-Naini}, \citenamefont {S{\"o}yler}, \citenamefont {Pupillo},
  \citenamefont {Sadeghpour},\ and\ \citenamefont
  {Capogrosso-Sansone}}]{sansone}%
  \BibitemOpen
  \bibfield  {author} {\bibinfo {author} {\bibfnamefont {A.}~\bibnamefont
  {Safavi-Naini}}, \bibinfo {author} {\bibfnamefont {S.~G.}\ \bibnamefont
  {S{\"o}yler}}, \bibinfo {author} {\bibfnamefont {G.}~\bibnamefont {Pupillo}},
  \bibinfo {author} {\bibfnamefont {H.~R.}\ \bibnamefont {Sadeghpour}}, \ and\
  \bibinfo {author} {\bibfnamefont {B.}~\bibnamefont {Capogrosso-Sansone}},\
  }\href {\doibase 10.1088/1367-2630/15/1/013036} {\ \textbf {\bibinfo {volume}
  {15}},\ \bibinfo {pages} {013036} (\bibinfo {year} {2013})}\BibitemShut
  {NoStop}%
\bibitem [{\citenamefont {Arg\"uelles}\ and\ \citenamefont
  {Santos}(2007)}]{santos_arturo}%
  \BibitemOpen
  \bibfield  {author} {\bibinfo {author} {\bibfnamefont {A.}~\bibnamefont
  {Arg\"uelles}}\ and\ \bibinfo {author} {\bibfnamefont {L.}~\bibnamefont
  {Santos}},\ }\href {\doibase 10.1103/PhysRevA.75.053613} {\bibfield
  {journal} {\bibinfo  {journal} {Phys. Rev. A}\ }\textbf {\bibinfo {volume}
  {75}},\ \bibinfo {pages} {053613} (\bibinfo {year} {2007})}\BibitemShut
  {NoStop}%
\bibitem [{\citenamefont {Singh}\ \emph {et~al.}(2017)\citenamefont {Singh},
  \citenamefont {Mondal}, \citenamefont {Sahoo},\ and\ \citenamefont
  {Mishra}}]{manpreet}%
  \BibitemOpen
  \bibfield  {author} {\bibinfo {author} {\bibfnamefont {M.}~\bibnamefont
  {Singh}}, \bibinfo {author} {\bibfnamefont {S.}~\bibnamefont {Mondal}},
  \bibinfo {author} {\bibfnamefont {B.~K.}\ \bibnamefont {Sahoo}}, \ and\
  \bibinfo {author} {\bibfnamefont {T.}~\bibnamefont {Mishra}},\ }\href
  {\doibase 10.1103/PhysRevA.96.053604} {\bibfield  {journal} {\bibinfo
  {journal} {Phys. Rev. A}\ }\textbf {\bibinfo {volume} {96}},\ \bibinfo
  {pages} {053604} (\bibinfo {year} {2017})}\BibitemShut {NoStop}%
\bibitem [{\citenamefont {Catani}\ \emph
  {et~al.}(2008{\natexlab{a}})\citenamefont {Catani}, \citenamefont {De~Sarlo},
  \citenamefont {Barontini}, \citenamefont {Minardi},\ and\ \citenamefont
  {Inguscio}}]{Inguscio}%
  \BibitemOpen
  \bibfield  {author} {\bibinfo {author} {\bibfnamefont {J.}~\bibnamefont
  {Catani}}, \bibinfo {author} {\bibfnamefont {L.}~\bibnamefont {De~Sarlo}},
  \bibinfo {author} {\bibfnamefont {G.}~\bibnamefont {Barontini}}, \bibinfo
  {author} {\bibfnamefont {F.}~\bibnamefont {Minardi}}, \ and\ \bibinfo
  {author} {\bibfnamefont {M.}~\bibnamefont {Inguscio}},\ }\href {\doibase
  10.1103/PhysRevA.77.011603} {\bibfield  {journal} {\bibinfo  {journal} {Phys.
  Rev. A}\ }\textbf {\bibinfo {volume} {77}},\ \bibinfo {pages} {011603}
  (\bibinfo {year} {2008}{\natexlab{a}})}\BibitemShut {NoStop}%
\bibitem [{\citenamefont {Hall}\ \emph {et~al.}(1998)\citenamefont {Hall},
  \citenamefont {Matthews}, \citenamefont {Ensher}, \citenamefont {Wieman},\
  and\ \citenamefont {Cornell}}]{Weiman}%
  \BibitemOpen
  \bibfield  {author} {\bibinfo {author} {\bibfnamefont {D.~S.}\ \bibnamefont
  {Hall}}, \bibinfo {author} {\bibfnamefont {M.~R.}\ \bibnamefont {Matthews}},
  \bibinfo {author} {\bibfnamefont {J.~R.}\ \bibnamefont {Ensher}}, \bibinfo
  {author} {\bibfnamefont {C.~E.}\ \bibnamefont {Wieman}}, \ and\ \bibinfo
  {author} {\bibfnamefont {E.~A.}\ \bibnamefont {Cornell}},\ }\href {\doibase
  10.1103/PhysRevLett.81.1539} {\bibfield  {journal} {\bibinfo  {journal}
  {Phys. Rev. Lett.}\ }\textbf {\bibinfo {volume} {81}},\ \bibinfo {pages}
  {1539} (\bibinfo {year} {1998})}\BibitemShut {NoStop}%
\bibitem [{\citenamefont {Catani}\ \emph
  {et~al.}(2008{\natexlab{b}})\citenamefont {Catani}, \citenamefont {De~Sarlo},
  \citenamefont {Barontini}, \citenamefont {Minardi},\ and\ \citenamefont
  {Inguscio}}]{catani1}%
  \BibitemOpen
  \bibfield  {author} {\bibinfo {author} {\bibfnamefont {J.}~\bibnamefont
  {Catani}}, \bibinfo {author} {\bibfnamefont {L.}~\bibnamefont {De~Sarlo}},
  \bibinfo {author} {\bibfnamefont {G.}~\bibnamefont {Barontini}}, \bibinfo
  {author} {\bibfnamefont {F.}~\bibnamefont {Minardi}}, \ and\ \bibinfo
  {author} {\bibfnamefont {M.}~\bibnamefont {Inguscio}},\ }\href {\doibase
  10.1103/PhysRevA.77.011603} {\bibfield  {journal} {\bibinfo  {journal} {Phys.
  Rev. A}\ }\textbf {\bibinfo {volume} {77}},\ \bibinfo {pages} {011603}
  (\bibinfo {year} {2008}{\natexlab{b}})}\BibitemShut {NoStop}%
\bibitem [{\citenamefont {Trotzky}\ \emph {et~al.}(2008)\citenamefont
  {Trotzky}, \citenamefont {Cheinet}, \citenamefont {F{\"o}lling},
  \citenamefont {Feld}, \citenamefont {Schnorrberger}, \citenamefont {Rey},
  \citenamefont {Polkovnikov}, \citenamefont {Demler}, \citenamefont {Lukin},\
  and\ \citenamefont {Bloch}}]{trotzky1}%
  \BibitemOpen
  \bibfield  {author} {\bibinfo {author} {\bibfnamefont {S.}~\bibnamefont
  {Trotzky}}, \bibinfo {author} {\bibfnamefont {P.}~\bibnamefont {Cheinet}},
  \bibinfo {author} {\bibfnamefont {S.}~\bibnamefont {F{\"o}lling}}, \bibinfo
  {author} {\bibfnamefont {M.}~\bibnamefont {Feld}}, \bibinfo {author}
  {\bibfnamefont {U.}~\bibnamefont {Schnorrberger}}, \bibinfo {author}
  {\bibfnamefont {A.~M.}\ \bibnamefont {Rey}}, \bibinfo {author} {\bibfnamefont
  {A.}~\bibnamefont {Polkovnikov}}, \bibinfo {author} {\bibfnamefont {E.~A.}\
  \bibnamefont {Demler}}, \bibinfo {author} {\bibfnamefont {M.~D.}\
  \bibnamefont {Lukin}}, \ and\ \bibinfo {author} {\bibfnamefont
  {I.}~\bibnamefont {Bloch}},\ }\href
  {https://science.sciencemag.org/content/319/5861/295} {\bibfield  {journal}
  {\bibinfo  {journal} {Science}\ }\textbf {\bibinfo {volume} {319}},\ \bibinfo
  {pages} {295} (\bibinfo {year} {2008})}\BibitemShut {NoStop}%
\bibitem [{\citenamefont {J{\"o}rdens}\ \emph {et~al.}(2008)\citenamefont
  {J{\"o}rdens}, \citenamefont {Strohmaier}, \citenamefont {G{\"u}nter},
  \citenamefont {Moritz},\ and\ \citenamefont {Esslinger}}]{jordens1}%
  \BibitemOpen
  \bibfield  {author} {\bibinfo {author} {\bibfnamefont {R.}~\bibnamefont
  {J{\"o}rdens}}, \bibinfo {author} {\bibfnamefont {N.}~\bibnamefont
  {Strohmaier}}, \bibinfo {author} {\bibfnamefont {K.}~\bibnamefont
  {G{\"u}nter}}, \bibinfo {author} {\bibfnamefont {H.}~\bibnamefont {Moritz}},
  \ and\ \bibinfo {author} {\bibfnamefont {T.}~\bibnamefont {Esslinger}},\
  }\href {https://www.nature.com/articles/nature07244} {\bibfield  {journal}
  {\bibinfo  {journal} {Nature}\ }\textbf {\bibinfo {volume} {455}},\ \bibinfo
  {pages} {204} (\bibinfo {year} {2008})}\BibitemShut {NoStop}%
\bibitem [{\citenamefont {Voigt}\ \emph {et~al.}(2009)\citenamefont {Voigt},
  \citenamefont {Taglieber}, \citenamefont {Costa}, \citenamefont {Aoki},
  \citenamefont {Wieser}, \citenamefont {H\"ansch},\ and\ \citenamefont
  {Dieckmann}}]{costa1}%
  \BibitemOpen
  \bibfield  {author} {\bibinfo {author} {\bibfnamefont {A.-C.}\ \bibnamefont
  {Voigt}}, \bibinfo {author} {\bibfnamefont {M.}~\bibnamefont {Taglieber}},
  \bibinfo {author} {\bibfnamefont {L.}~\bibnamefont {Costa}}, \bibinfo
  {author} {\bibfnamefont {T.}~\bibnamefont {Aoki}}, \bibinfo {author}
  {\bibfnamefont {W.}~\bibnamefont {Wieser}}, \bibinfo {author} {\bibfnamefont
  {T.~W.}\ \bibnamefont {H\"ansch}}, \ and\ \bibinfo {author} {\bibfnamefont
  {K.}~\bibnamefont {Dieckmann}},\ }\href {\doibase
  10.1103/PhysRevLett.102.020405} {\bibfield  {journal} {\bibinfo  {journal}
  {Phys. Rev. Lett.}\ }\textbf {\bibinfo {volume} {102}},\ \bibinfo {pages}
  {020405} (\bibinfo {year} {2009})}\BibitemShut {NoStop}%
\bibitem [{\citenamefont {Snoek}\ \emph {et~al.}(2011)\citenamefont {Snoek},
  \citenamefont {Titvinidze}, \citenamefont {Bloch},\ and\ \citenamefont
  {Hofstetter}}]{bloch2011}%
  \BibitemOpen
  \bibfield  {author} {\bibinfo {author} {\bibfnamefont {M.}~\bibnamefont
  {Snoek}}, \bibinfo {author} {\bibfnamefont {I.}~\bibnamefont {Titvinidze}},
  \bibinfo {author} {\bibfnamefont {I.}~\bibnamefont {Bloch}}, \ and\ \bibinfo
  {author} {\bibfnamefont {W.}~\bibnamefont {Hofstetter}},\ }\href {\doibase
  10.1103/PhysRevLett.106.155301} {\bibfield  {journal} {\bibinfo  {journal}
  {Phys. Rev. Lett.}\ }\textbf {\bibinfo {volume} {106}},\ \bibinfo {pages}
  {155301} (\bibinfo {year} {2011})}\BibitemShut {NoStop}%
\bibitem [{\citenamefont {G\"unter}\ \emph {et~al.}(2006)\citenamefont
  {G\"unter}, \citenamefont {St\"oferle}, \citenamefont {Moritz}, \citenamefont
  {K\"ohl},\ and\ \citenamefont {Esslinger}}]{moritz1}%
  \BibitemOpen
  \bibfield  {author} {\bibinfo {author} {\bibfnamefont {K.}~\bibnamefont
  {G\"unter}}, \bibinfo {author} {\bibfnamefont {T.}~\bibnamefont
  {St\"oferle}}, \bibinfo {author} {\bibfnamefont {H.}~\bibnamefont {Moritz}},
  \bibinfo {author} {\bibfnamefont {M.}~\bibnamefont {K\"ohl}}, \ and\ \bibinfo
  {author} {\bibfnamefont {T.}~\bibnamefont {Esslinger}},\ }\href {\doibase
  10.1103/PhysRevLett.96.180402} {\bibfield  {journal} {\bibinfo  {journal}
  {Phys. Rev. Lett.}\ }\textbf {\bibinfo {volume} {96}},\ \bibinfo {pages}
  {180402} (\bibinfo {year} {2006})}\BibitemShut {NoStop}%
\bibitem [{\citenamefont {Roati}\ \emph {et~al.}(2002)\citenamefont {Roati},
  \citenamefont {Riboli}, \citenamefont {Modugno},\ and\ \citenamefont
  {Inguscio}}]{roati1}%
  \BibitemOpen
  \bibfield  {author} {\bibinfo {author} {\bibfnamefont {G.}~\bibnamefont
  {Roati}}, \bibinfo {author} {\bibfnamefont {F.}~\bibnamefont {Riboli}},
  \bibinfo {author} {\bibfnamefont {G.}~\bibnamefont {Modugno}}, \ and\
  \bibinfo {author} {\bibfnamefont {M.}~\bibnamefont {Inguscio}},\ }\href
  {\doibase 10.1103/PhysRevLett.89.150403} {\bibfield  {journal} {\bibinfo
  {journal} {Phys. Rev. Lett.}\ }\textbf {\bibinfo {volume} {89}},\ \bibinfo
  {pages} {150403} (\bibinfo {year} {2002})}\BibitemShut {NoStop}%
\bibitem [{\citenamefont {Best}\ \emph {et~al.}(2009)\citenamefont {Best},
  \citenamefont {Will}, \citenamefont {Schneider}, \citenamefont
  {Hackerm\"uller}, \citenamefont {van Oosten}, \citenamefont {Bloch},\ and\
  \citenamefont {L\"uhmann}}]{best1}%
  \BibitemOpen
  \bibfield  {author} {\bibinfo {author} {\bibfnamefont {T.}~\bibnamefont
  {Best}}, \bibinfo {author} {\bibfnamefont {S.}~\bibnamefont {Will}}, \bibinfo
  {author} {\bibfnamefont {U.}~\bibnamefont {Schneider}}, \bibinfo {author}
  {\bibfnamefont {L.}~\bibnamefont {Hackerm\"uller}}, \bibinfo {author}
  {\bibfnamefont {D.}~\bibnamefont {van Oosten}}, \bibinfo {author}
  {\bibfnamefont {I.}~\bibnamefont {Bloch}}, \ and\ \bibinfo {author}
  {\bibfnamefont {D.-S.}\ \bibnamefont {L\"uhmann}},\ }\href {\doibase
  10.1103/PhysRevLett.102.030408} {\bibfield  {journal} {\bibinfo  {journal}
  {Phys. Rev. Lett.}\ }\textbf {\bibinfo {volume} {102}},\ \bibinfo {pages}
  {030408} (\bibinfo {year} {2009})}\BibitemShut {NoStop}%
\bibitem [{\citenamefont {He}\ \emph {et~al.}(2012)\citenamefont {He},
  \citenamefont {Li}, \citenamefont {Altman},\ and\ \citenamefont
  {Hofstetter}}]{altman}%
  \BibitemOpen
  \bibfield  {author} {\bibinfo {author} {\bibfnamefont {L.}~\bibnamefont
  {He}}, \bibinfo {author} {\bibfnamefont {Y.}~\bibnamefont {Li}}, \bibinfo
  {author} {\bibfnamefont {E.}~\bibnamefont {Altman}}, \ and\ \bibinfo {author}
  {\bibfnamefont {W.}~\bibnamefont {Hofstetter}},\ }\href {\doibase
  10.1103/PhysRevA.86.043620} {\bibfield  {journal} {\bibinfo  {journal} {Phys.
  Rev. A}\ }\textbf {\bibinfo {volume} {86}},\ \bibinfo {pages} {043620}
  (\bibinfo {year} {2012})}\BibitemShut {NoStop}%
\bibitem [{\citenamefont {H\'ebert}\ \emph {et~al.}(2008)\citenamefont
  {H\'ebert}, \citenamefont {Batrouni}, \citenamefont {Roy},\ and\
  \citenamefont {Rousseau}}]{batrouni}%
  \BibitemOpen
  \bibfield  {author} {\bibinfo {author} {\bibfnamefont {F.}~\bibnamefont
  {H\'ebert}}, \bibinfo {author} {\bibfnamefont {G.~G.}\ \bibnamefont
  {Batrouni}}, \bibinfo {author} {\bibfnamefont {X.}~\bibnamefont {Roy}}, \
  and\ \bibinfo {author} {\bibfnamefont {V.~G.}\ \bibnamefont {Rousseau}},\
  }\href {\doibase 10.1103/PhysRevB.78.184505} {\bibfield  {journal} {\bibinfo
  {journal} {Phys. Rev. B}\ }\textbf {\bibinfo {volume} {78}},\ \bibinfo
  {pages} {184505} (\bibinfo {year} {2008})}\BibitemShut {NoStop}%
\bibitem [{\citenamefont {Mishra}\ \emph {et~al.}(2007)\citenamefont {Mishra},
  \citenamefont {Pai},\ and\ \citenamefont {Das}}]{mishra1}%
  \BibitemOpen
  \bibfield  {author} {\bibinfo {author} {\bibfnamefont {T.}~\bibnamefont
  {Mishra}}, \bibinfo {author} {\bibfnamefont {R.~V.}\ \bibnamefont {Pai}}, \
  and\ \bibinfo {author} {\bibfnamefont {B.~P.}\ \bibnamefont {Das}},\ }\href
  {\doibase 10.1103/PhysRevA.76.013604} {\bibfield  {journal} {\bibinfo
  {journal} {Phys. Rev. A}\ }\textbf {\bibinfo {volume} {76}},\ \bibinfo
  {pages} {013604} (\bibinfo {year} {2007})}\BibitemShut {NoStop}%
\bibitem [{\citenamefont {Mishra}\ \emph {et~al.}(2008)\citenamefont {Mishra},
  \citenamefont {Sahoo},\ and\ \citenamefont {Pai}}]{mishra2}%
  \BibitemOpen
  \bibfield  {author} {\bibinfo {author} {\bibfnamefont {T.}~\bibnamefont
  {Mishra}}, \bibinfo {author} {\bibfnamefont {B.~K.}\ \bibnamefont {Sahoo}}, \
  and\ \bibinfo {author} {\bibfnamefont {R.~V.}\ \bibnamefont {Pai}},\ }\href
  {\doibase 10.1103/PhysRevA.78.013632} {\bibfield  {journal} {\bibinfo
  {journal} {Phys. Rev. A}\ }\textbf {\bibinfo {volume} {78}},\ \bibinfo
  {pages} {013632} (\bibinfo {year} {2008})}\BibitemShut {NoStop}%
\bibitem [{\citenamefont {Ozaki}\ and\ \citenamefont {Nikuni}(2012)}]{ozaki1}%
  \BibitemOpen
  \bibfield  {author} {\bibinfo {author} {\bibfnamefont {T.}~\bibnamefont
  {Ozaki}}\ and\ \bibinfo {author} {\bibfnamefont {T.}~\bibnamefont {Nikuni}},\
  }\href {\doibase 10.1143/JPSJ.81.024001} {\bibfield  {journal} {\bibinfo
  {journal} {Journal of the Physical Society of Japan}\ }\textbf {\bibinfo
  {volume} {81}},\ \bibinfo {pages} {024001} (\bibinfo {year}
  {2012})}\BibitemShut {NoStop}%
\bibitem [{\citenamefont {Pu}\ and\ \citenamefont {Bigelow}(1998)}]{pu1}%
  \BibitemOpen
  \bibfield  {author} {\bibinfo {author} {\bibfnamefont {H.}~\bibnamefont
  {Pu}}\ and\ \bibinfo {author} {\bibfnamefont {N.~P.}\ \bibnamefont
  {Bigelow}},\ }\href {\doibase 10.1103/PhysRevLett.80.1130} {\bibfield
  {journal} {\bibinfo  {journal} {Phys. Rev. Lett.}\ }\textbf {\bibinfo
  {volume} {80}},\ \bibinfo {pages} {1130} (\bibinfo {year}
  {1998})}\BibitemShut {NoStop}%
\bibitem [{\citenamefont {Kuklov}\ \emph {et~al.}(2004)\citenamefont {Kuklov},
  \citenamefont {Prokof'ev},\ and\ \citenamefont {Svistunov}}]{kuklov1}%
  \BibitemOpen
  \bibfield  {author} {\bibinfo {author} {\bibfnamefont {A.}~\bibnamefont
  {Kuklov}}, \bibinfo {author} {\bibfnamefont {N.}~\bibnamefont {Prokof'ev}}, \
  and\ \bibinfo {author} {\bibfnamefont {B.}~\bibnamefont {Svistunov}},\ }\href
  {\doibase 10.1103/PhysRevLett.92.050402} {\bibfield  {journal} {\bibinfo
  {journal} {Phys. Rev. Lett.}\ }\textbf {\bibinfo {volume} {92}},\ \bibinfo
  {pages} {050402} (\bibinfo {year} {2004})}\BibitemShut {NoStop}%
\bibitem [{\citenamefont {Mathey}(2007)}]{mathey1}%
  \BibitemOpen
  \bibfield  {author} {\bibinfo {author} {\bibfnamefont {L.}~\bibnamefont
  {Mathey}},\ }\href {\doibase 10.1103/PhysRevB.75.144510} {\bibfield
  {journal} {\bibinfo  {journal} {Phys. Rev. B}\ }\textbf {\bibinfo {volume}
  {75}},\ \bibinfo {pages} {144510} (\bibinfo {year} {2007})}\BibitemShut
  {NoStop}%
\bibitem [{\citenamefont {Isacsson}\ \emph {et~al.}(2005)\citenamefont
  {Isacsson}, \citenamefont {Cha}, \citenamefont {Sengupta},\ and\
  \citenamefont {Girvin}}]{isacsson1}%
  \BibitemOpen
  \bibfield  {author} {\bibinfo {author} {\bibfnamefont {A.}~\bibnamefont
  {Isacsson}}, \bibinfo {author} {\bibfnamefont {M.-C.}\ \bibnamefont {Cha}},
  \bibinfo {author} {\bibfnamefont {K.}~\bibnamefont {Sengupta}}, \ and\
  \bibinfo {author} {\bibfnamefont {S.~M.}\ \bibnamefont {Girvin}},\ }\href
  {\doibase 10.1103/PhysRevB.72.184507} {\bibfield  {journal} {\bibinfo
  {journal} {Phys. Rev. B}\ }\textbf {\bibinfo {volume} {72}},\ \bibinfo
  {pages} {184507} (\bibinfo {year} {2005})}\BibitemShut {NoStop}%
\bibitem [{\citenamefont {{S Sugawa}}\ \emph {et~al.}(2011)\citenamefont {{S
  Sugawa}}, \citenamefont {{K Inaba}}, \citenamefont {{S Taie}}, \citenamefont
  {{R Yamazaki}}, \citenamefont {{M Yamashita}},\ and\ \citenamefont {{Y
  Takahashi}}}]{Sugawa2011}%
  \BibitemOpen
  \bibfield  {author} {\bibinfo {author} {\bibnamefont {{S Sugawa}}}, \bibinfo
  {author} {\bibnamefont {{K Inaba}}}, \bibinfo {author} {\bibnamefont {{S
  Taie}}}, \bibinfo {author} {\bibnamefont {{R Yamazaki}}}, \bibinfo {author}
  {\bibnamefont {{M Yamashita}}}, \ and\ \bibinfo {author} {\bibnamefont {{Y
  Takahashi}}},\ }\href {\doibase https://doi.org/10.1038/nphys2028} {\bibfield
   {journal} {\bibinfo  {journal} {Nature Physics}\ }\textbf {\bibinfo {volume}
  {7}},\ \bibinfo {pages} {642} (\bibinfo {year} {2011})}\BibitemShut {NoStop}%
\bibitem [{\citenamefont {Hara}\ \emph {et~al.}(2014)\citenamefont {Hara},
  \citenamefont {Konishi}, \citenamefont {Nakajima}, \citenamefont {Takasu},\
  and\ \citenamefont {Takahashi}}]{takahashi}%
  \BibitemOpen
  \bibfield  {author} {\bibinfo {author} {\bibfnamefont {H.}~\bibnamefont
  {Hara}}, \bibinfo {author} {\bibfnamefont {H.}~\bibnamefont {Konishi}},
  \bibinfo {author} {\bibfnamefont {S.}~\bibnamefont {Nakajima}}, \bibinfo
  {author} {\bibfnamefont {Y.}~\bibnamefont {Takasu}}, \ and\ \bibinfo {author}
  {\bibfnamefont {Y.}~\bibnamefont {Takahashi}},\ }\href {\doibase
  10.7566/JPSJ.83.014003} {\bibfield  {journal} {\bibinfo  {journal} {Journal
  of the Physical Society of Japan}\ }\textbf {\bibinfo {volume} {83}},\
  \bibinfo {pages} {014003} (\bibinfo {year} {2014})}\BibitemShut {NoStop}%
\bibitem [{\citenamefont {Peil}\ \emph {et~al.}(2003)\citenamefont {Peil},
  \citenamefont {Porto}, \citenamefont {Tolra}, \citenamefont {Obrecht},
  \citenamefont {King}, \citenamefont {Subbotin}, \citenamefont {Rolston},\
  and\ \citenamefont {Phillips}}]{Porto1}%
  \BibitemOpen
  \bibfield  {author} {\bibinfo {author} {\bibfnamefont {S.}~\bibnamefont
  {Peil}}, \bibinfo {author} {\bibfnamefont {J.~V.}\ \bibnamefont {Porto}},
  \bibinfo {author} {\bibfnamefont {B.~L.}\ \bibnamefont {Tolra}}, \bibinfo
  {author} {\bibfnamefont {J.~M.}\ \bibnamefont {Obrecht}}, \bibinfo {author}
  {\bibfnamefont {B.~E.}\ \bibnamefont {King}}, \bibinfo {author}
  {\bibfnamefont {M.}~\bibnamefont {Subbotin}}, \bibinfo {author}
  {\bibfnamefont {S.~L.}\ \bibnamefont {Rolston}}, \ and\ \bibinfo {author}
  {\bibfnamefont {W.~D.}\ \bibnamefont {Phillips}},\ }\href {\doibase
  10.1103/PhysRevA.67.051603} {\bibfield  {journal} {\bibinfo  {journal} {Phys.
  Rev. A}\ }\textbf {\bibinfo {volume} {67}},\ \bibinfo {pages} {051603}
  (\bibinfo {year} {2003})}\BibitemShut {NoStop}%
\bibitem [{\citenamefont {Sebby-Strabley}\ \emph
  {et~al.}(2006{\natexlab{a}})\citenamefont {Sebby-Strabley}, \citenamefont
  {Anderlini}, \citenamefont {Jessen},\ and\ \citenamefont {Porto}}]{Porto2}%
  \BibitemOpen
  \bibfield  {author} {\bibinfo {author} {\bibfnamefont {J.}~\bibnamefont
  {Sebby-Strabley}}, \bibinfo {author} {\bibfnamefont {M.}~\bibnamefont
  {Anderlini}}, \bibinfo {author} {\bibfnamefont {P.~S.}\ \bibnamefont
  {Jessen}}, \ and\ \bibinfo {author} {\bibfnamefont {J.~V.}\ \bibnamefont
  {Porto}},\ }\href {\doibase 10.1103/PhysRevA.73.033605} {\bibfield  {journal}
  {\bibinfo  {journal} {Phys. Rev. A}\ }\textbf {\bibinfo {volume} {73}},\
  \bibinfo {pages} {033605} (\bibinfo {year} {2006}{\natexlab{a}})}\BibitemShut
  {NoStop}%
\bibitem [{\citenamefont {Singh}\ \emph
  {et~al.}(2012{\natexlab{b}})\citenamefont {Singh}, \citenamefont {Dhar},
  \citenamefont {Mishra}, \citenamefont {Pai},\ and\ \citenamefont
  {Das}}]{manpreetsuperlat1}%
  \BibitemOpen
  \bibfield  {author} {\bibinfo {author} {\bibfnamefont {M.}~\bibnamefont
  {Singh}}, \bibinfo {author} {\bibfnamefont {A.}~\bibnamefont {Dhar}},
  \bibinfo {author} {\bibfnamefont {T.}~\bibnamefont {Mishra}}, \bibinfo
  {author} {\bibfnamefont {R.~V.}\ \bibnamefont {Pai}}, \ and\ \bibinfo
  {author} {\bibfnamefont {B.~P.}\ \bibnamefont {Das}},\ }\href {\doibase
  10.1103/PhysRevA.85.051604} {\bibfield  {journal} {\bibinfo  {journal} {Phys.
  Rev. A}\ }\textbf {\bibinfo {volume} {85}},\ \bibinfo {pages} {051604}
  (\bibinfo {year} {2012}{\natexlab{b}})}\BibitemShut {NoStop}%
\bibitem [{\citenamefont {Singh}\ and\ \citenamefont
  {Mishra}(2016)}]{manpreetsuperlat2}%
  \BibitemOpen
  \bibfield  {author} {\bibinfo {author} {\bibfnamefont {M.}~\bibnamefont
  {Singh}}\ and\ \bibinfo {author} {\bibfnamefont {T.}~\bibnamefont {Mishra}},\
  }\href {\doibase 10.1103/PhysRevA.94.063610} {\bibfield  {journal} {\bibinfo
  {journal} {Phys. Rev. A}\ }\textbf {\bibinfo {volume} {94}},\ \bibinfo
  {pages} {063610} (\bibinfo {year} {2016})}\BibitemShut {NoStop}%
\bibitem [{\citenamefont {Dhar}\ \emph
  {et~al.}(2011{\natexlab{a}})\citenamefont {Dhar}, \citenamefont {Mishra},
  \citenamefont {Pai},\ and\ \citenamefont {Das}}]{aryasuperlat1}%
  \BibitemOpen
  \bibfield  {author} {\bibinfo {author} {\bibfnamefont {A.}~\bibnamefont
  {Dhar}}, \bibinfo {author} {\bibfnamefont {T.}~\bibnamefont {Mishra}},
  \bibinfo {author} {\bibfnamefont {R.~V.}\ \bibnamefont {Pai}}, \ and\
  \bibinfo {author} {\bibfnamefont {B.~P.}\ \bibnamefont {Das}},\ }\href
  {\doibase 10.1103/PhysRevA.83.053621} {\bibfield  {journal} {\bibinfo
  {journal} {Phys. Rev. A}\ }\textbf {\bibinfo {volume} {83}},\ \bibinfo
  {pages} {053621} (\bibinfo {year} {2011}{\natexlab{a}})}\BibitemShut
  {NoStop}%
\bibitem [{\citenamefont {Dhar}\ \emph
  {et~al.}(2011{\natexlab{b}})\citenamefont {Dhar}, \citenamefont {Singh},
  \citenamefont {Pai},\ and\ \citenamefont {Das}}]{aryasuperlat2}%
  \BibitemOpen
  \bibfield  {author} {\bibinfo {author} {\bibfnamefont {A.}~\bibnamefont
  {Dhar}}, \bibinfo {author} {\bibfnamefont {M.}~\bibnamefont {Singh}},
  \bibinfo {author} {\bibfnamefont {R.~V.}\ \bibnamefont {Pai}}, \ and\
  \bibinfo {author} {\bibfnamefont {B.~P.}\ \bibnamefont {Das}},\ }\href
  {\doibase 10.1103/PhysRevA.84.033631} {\bibfield  {journal} {\bibinfo
  {journal} {Phys. Rev. A}\ }\textbf {\bibinfo {volume} {84}},\ \bibinfo
  {pages} {033631} (\bibinfo {year} {2011}{\natexlab{b}})}\BibitemShut
  {NoStop}%
\bibitem [{\citenamefont {Dhar}\ \emph {et~al.}(2013)\citenamefont {Dhar},
  \citenamefont {Mishra}, \citenamefont {Pai}, \citenamefont {Mukerjee},\ and\
  \citenamefont {Das}}]{aryasuperlat3}%
  \BibitemOpen
  \bibfield  {author} {\bibinfo {author} {\bibfnamefont {A.}~\bibnamefont
  {Dhar}}, \bibinfo {author} {\bibfnamefont {T.}~\bibnamefont {Mishra}},
  \bibinfo {author} {\bibfnamefont {R.~V.}\ \bibnamefont {Pai}}, \bibinfo
  {author} {\bibfnamefont {S.}~\bibnamefont {Mukerjee}}, \ and\ \bibinfo
  {author} {\bibfnamefont {B.~P.}\ \bibnamefont {Das}},\ }\href {\doibase
  10.1103/PhysRevA.88.053625} {\bibfield  {journal} {\bibinfo  {journal} {Phys.
  Rev. A}\ }\textbf {\bibinfo {volume} {88}},\ \bibinfo {pages} {053625}
  (\bibinfo {year} {2013})}\BibitemShut {NoStop}%
\bibitem [{\citenamefont {Roth}\ and\ \citenamefont {Burnett}(2003)}]{roth1}%
  \BibitemOpen
  \bibfield  {author} {\bibinfo {author} {\bibfnamefont {R.}~\bibnamefont
  {Roth}}\ and\ \bibinfo {author} {\bibfnamefont {K.}~\bibnamefont {Burnett}},\
  }\href {\doibase 10.1103/PhysRevA.68.023604} {\bibfield  {journal} {\bibinfo
  {journal} {Phys. Rev. A}\ }\textbf {\bibinfo {volume} {68}},\ \bibinfo
  {pages} {023604} (\bibinfo {year} {2003})}\BibitemShut {NoStop}%
\bibitem [{\citenamefont {Schmitt}\ \emph {et~al.}(2009)\citenamefont
  {Schmitt}, \citenamefont {Hild},\ and\ \citenamefont {Roth}}]{roth2}%
  \BibitemOpen
  \bibfield  {author} {\bibinfo {author} {\bibfnamefont {F.}~\bibnamefont
  {Schmitt}}, \bibinfo {author} {\bibfnamefont {M.}~\bibnamefont {Hild}}, \
  and\ \bibinfo {author} {\bibfnamefont {R.}~\bibnamefont {Roth}},\ }\href
  {\doibase 10.1103/PhysRevA.80.023621} {\bibfield  {journal} {\bibinfo
  {journal} {Phys. Rev. A}\ }\textbf {\bibinfo {volume} {80}},\ \bibinfo
  {pages} {023621} (\bibinfo {year} {2009})}\BibitemShut {NoStop}%
\bibitem [{\citenamefont {Roux}\ \emph {et~al.}(2008)\citenamefont {Roux},
  \citenamefont {Barthel}, \citenamefont {McCulloch}, \citenamefont {Kollath},
  \citenamefont {Schollw\"ock},\ and\ \citenamefont {Giamarchi}}]{roux1}%
  \BibitemOpen
  \bibfield  {author} {\bibinfo {author} {\bibfnamefont {G.}~\bibnamefont
  {Roux}}, \bibinfo {author} {\bibfnamefont {T.}~\bibnamefont {Barthel}},
  \bibinfo {author} {\bibfnamefont {I.~P.}\ \bibnamefont {McCulloch}}, \bibinfo
  {author} {\bibfnamefont {C.}~\bibnamefont {Kollath}}, \bibinfo {author}
  {\bibfnamefont {U.}~\bibnamefont {Schollw\"ock}}, \ and\ \bibinfo {author}
  {\bibfnamefont {T.}~\bibnamefont {Giamarchi}},\ }\href {\doibase
  10.1103/PhysRevA.78.023628} {\bibfield  {journal} {\bibinfo  {journal} {Phys.
  Rev. A}\ }\textbf {\bibinfo {volume} {78}},\ \bibinfo {pages} {023628}
  (\bibinfo {year} {2008})}\BibitemShut {NoStop}%
\bibitem [{\citenamefont {Sebby-Strabley}\ \emph
  {et~al.}(2006{\natexlab{b}})\citenamefont {Sebby-Strabley}, \citenamefont
  {Anderlini}, \citenamefont {Jessen},\ and\ \citenamefont {Porto}}]{sebby1}%
  \BibitemOpen
  \bibfield  {author} {\bibinfo {author} {\bibfnamefont {J.}~\bibnamefont
  {Sebby-Strabley}}, \bibinfo {author} {\bibfnamefont {M.}~\bibnamefont
  {Anderlini}}, \bibinfo {author} {\bibfnamefont {P.~S.}\ \bibnamefont
  {Jessen}}, \ and\ \bibinfo {author} {\bibfnamefont {J.~V.}\ \bibnamefont
  {Porto}},\ }\href {\doibase 10.1103/PhysRevA.73.033605} {\bibfield  {journal}
  {\bibinfo  {journal} {Phys. Rev. A}\ }\textbf {\bibinfo {volume} {73}},\
  \bibinfo {pages} {033605} (\bibinfo {year} {2006}{\natexlab{b}})}\BibitemShut
  {NoStop}%
\bibitem [{\citenamefont {Cheinet}\ \emph {et~al.}(2008)\citenamefont
  {Cheinet}, \citenamefont {Trotzky}, \citenamefont {Feld}, \citenamefont
  {Schnorrberger}, \citenamefont {Moreno-Cardoner}, \citenamefont {F\"olling},\
  and\ \citenamefont {Bloch}}]{cheinet1}%
  \BibitemOpen
  \bibfield  {author} {\bibinfo {author} {\bibfnamefont {P.}~\bibnamefont
  {Cheinet}}, \bibinfo {author} {\bibfnamefont {S.}~\bibnamefont {Trotzky}},
  \bibinfo {author} {\bibfnamefont {M.}~\bibnamefont {Feld}}, \bibinfo {author}
  {\bibfnamefont {U.}~\bibnamefont {Schnorrberger}}, \bibinfo {author}
  {\bibfnamefont {M.}~\bibnamefont {Moreno-Cardoner}}, \bibinfo {author}
  {\bibfnamefont {S.}~\bibnamefont {F\"olling}}, \ and\ \bibinfo {author}
  {\bibfnamefont {I.}~\bibnamefont {Bloch}},\ }\href {\doibase
  10.1103/PhysRevLett.101.090404} {\bibfield  {journal} {\bibinfo  {journal}
  {Phys. Rev. Lett.}\ }\textbf {\bibinfo {volume} {101}},\ \bibinfo {pages}
  {090404} (\bibinfo {year} {2008})}\BibitemShut {NoStop}%
\bibitem [{\citenamefont {Grusdt}\ \emph {et~al.}(2013)\citenamefont {Grusdt},
  \citenamefont {H\"oning},\ and\ \citenamefont {Fleischhauer}}]{fleischhauer}%
  \BibitemOpen
  \bibfield  {author} {\bibinfo {author} {\bibfnamefont {F.}~\bibnamefont
  {Grusdt}}, \bibinfo {author} {\bibfnamefont {M.}~\bibnamefont {H\"oning}}, \
  and\ \bibinfo {author} {\bibfnamefont {M.}~\bibnamefont {Fleischhauer}},\
  }\href {\doibase 10.1103/PhysRevLett.110.260405} {\bibfield  {journal}
  {\bibinfo  {journal} {Phys. Rev. Lett.}\ }\textbf {\bibinfo {volume} {110}},\
  \bibinfo {pages} {260405} (\bibinfo {year} {2013})}\BibitemShut {NoStop}%
\bibitem [{\citenamefont {Nascimb\`ene}\ \emph {et~al.}(2012)\citenamefont
  {Nascimb\`ene}, \citenamefont {Chen}, \citenamefont {Atala}, \citenamefont
  {Aidelsburger}, \citenamefont {Trotzky}, \citenamefont {Paredes},\ and\
  \citenamefont {Bloch}}]{Atala2013}%
  \BibitemOpen
  \bibfield  {author} {\bibinfo {author} {\bibfnamefont {S.}~\bibnamefont
  {Nascimb\`ene}}, \bibinfo {author} {\bibfnamefont {Y.-A.}\ \bibnamefont
  {Chen}}, \bibinfo {author} {\bibfnamefont {M.}~\bibnamefont {Atala}},
  \bibinfo {author} {\bibfnamefont {M.}~\bibnamefont {Aidelsburger}}, \bibinfo
  {author} {\bibfnamefont {S.}~\bibnamefont {Trotzky}}, \bibinfo {author}
  {\bibfnamefont {B.}~\bibnamefont {Paredes}}, \ and\ \bibinfo {author}
  {\bibfnamefont {I.}~\bibnamefont {Bloch}},\ }\href {\doibase
  10.1103/PhysRevLett.108.205301} {\bibfield  {journal} {\bibinfo  {journal}
  {Phys. Rev. Lett.}\ }\textbf {\bibinfo {volume} {108}},\ \bibinfo {pages}
  {205301} (\bibinfo {year} {2012})}\BibitemShut {NoStop}%
\bibitem [{\citenamefont {Altman}\ \emph {et~al.}(2003)\citenamefont {Altman},
  \citenamefont {Hofstetter}, \citenamefont {Demler},\ and\ \citenamefont
  {Lukin}}]{Altman_2003}%
  \BibitemOpen
  \bibfield  {author} {\bibinfo {author} {\bibfnamefont {E.}~\bibnamefont
  {Altman}}, \bibinfo {author} {\bibfnamefont {W.}~\bibnamefont {Hofstetter}},
  \bibinfo {author} {\bibfnamefont {E.}~\bibnamefont {Demler}}, \ and\ \bibinfo
  {author} {\bibfnamefont {M.~D.}\ \bibnamefont {Lukin}},\ }\href {\doibase
  10.1088/1367-2630/5/1/113} {\bibfield  {journal} {\bibinfo  {journal} {New
  Journal of Physics}\ }\textbf {\bibinfo {volume} {5}},\ \bibinfo {pages}
  {113–113} (\bibinfo {year} {2003})}\BibitemShut {NoStop}%
\bibitem [{\citenamefont {Chen}\ \emph {et~al.}(2010)\citenamefont {Chen},
  \citenamefont {Kou}, \citenamefont {Zhang},\ and\ \citenamefont
  {Chen}}]{shuchen1}%
  \BibitemOpen
  \bibfield  {author} {\bibinfo {author} {\bibfnamefont {B.-L.}\ \bibnamefont
  {Chen}}, \bibinfo {author} {\bibfnamefont {S.-P.}\ \bibnamefont {Kou}},
  \bibinfo {author} {\bibfnamefont {Y.}~\bibnamefont {Zhang}}, \ and\ \bibinfo
  {author} {\bibfnamefont {S.}~\bibnamefont {Chen}},\ }\href {\doibase
  10.1103/PhysRevA.81.053608} {\bibfield  {journal} {\bibinfo  {journal} {Phys.
  Rev. A}\ }\textbf {\bibinfo {volume} {81}},\ \bibinfo {pages} {053608}
  (\bibinfo {year} {2010})}\BibitemShut {NoStop}%
\bibitem [{\citenamefont {Sheshadri}\ \emph {et~al.}(1993)\citenamefont
  {Sheshadri}, \citenamefont {Krishnamurthy}, \citenamefont {Pandit},\ and\
  \citenamefont {Ramakrishnan}}]{sheshadri1}%
  \BibitemOpen
  \bibfield  {author} {\bibinfo {author} {\bibfnamefont {K.}~\bibnamefont
  {Sheshadri}}, \bibinfo {author} {\bibfnamefont {H.}~\bibnamefont
  {Krishnamurthy}}, \bibinfo {author} {\bibfnamefont {R.}~\bibnamefont
  {Pandit}}, \ and\ \bibinfo {author} {\bibfnamefont {T.}~\bibnamefont
  {Ramakrishnan}},\ }\href
  {https://iopscience.iop.org/article/10.1209/0295-5075/22/4/004/meta}
  {\bibfield  {journal} {\bibinfo  {journal} {EPL (Europhysics Letters)}\
  }\textbf {\bibinfo {volume} {22}},\ \bibinfo {pages} {257} (\bibinfo {year}
  {1993})}\BibitemShut {NoStop}%
\bibitem [{\citenamefont {Danshita}\ \emph {et~al.}(2007)\citenamefont
  {Danshita}, \citenamefont {Williams}, \citenamefont {S\'a~de Melo},\ and\
  \citenamefont {Clark}}]{danshita8}%
  \BibitemOpen
  \bibfield  {author} {\bibinfo {author} {\bibfnamefont {I.}~\bibnamefont
  {Danshita}}, \bibinfo {author} {\bibfnamefont {J.~E.}\ \bibnamefont
  {Williams}}, \bibinfo {author} {\bibfnamefont {C.~A.~R.}\ \bibnamefont
  {S\'a~de Melo}}, \ and\ \bibinfo {author} {\bibfnamefont {C.~W.}\
  \bibnamefont {Clark}},\ }\href {\doibase 10.1103/PhysRevA.76.043606}
  {\bibfield  {journal} {\bibinfo  {journal} {Phys. Rev. A}\ }\textbf {\bibinfo
  {volume} {76}},\ \bibinfo {pages} {043606} (\bibinfo {year}
  {2007})}\BibitemShut {NoStop}%
\bibitem [{\citenamefont {Chen}\ and\ \citenamefont
  {Yang}(2017)}]{chen2017two}%
  \BibitemOpen
  \bibfield  {author} {\bibinfo {author} {\bibfnamefont {Y.-C.}\ \bibnamefont
  {Chen}}\ and\ \bibinfo {author} {\bibfnamefont {M.-F.}\ \bibnamefont
  {Yang}},\ }\href
  {https://iopscience.iop.org/article/10.1088/2399-6528/aa8bfb/meta} {\bibfield
   {journal} {\bibinfo  {journal} {Journal of Physics Communications}\ }\textbf
  {\bibinfo {volume} {1}},\ \bibinfo {pages} {035009} (\bibinfo {year}
  {2017})}\BibitemShut {NoStop}%
\bibitem [{\citenamefont {Yang}\ \emph {et~al.}(2017)\citenamefont {Yang},
  \citenamefont {Dai}, \citenamefont {Sun}, \citenamefont {Reingruber},
  \citenamefont {Yuan},\ and\ \citenamefont {Pan}}]{PhysRevA.96.011602}%
  \BibitemOpen
  \bibfield  {author} {\bibinfo {author} {\bibfnamefont {B.}~\bibnamefont
  {Yang}}, \bibinfo {author} {\bibfnamefont {H.-N.}\ \bibnamefont {Dai}},
  \bibinfo {author} {\bibfnamefont {H.}~\bibnamefont {Sun}}, \bibinfo {author}
  {\bibfnamefont {A.}~\bibnamefont {Reingruber}}, \bibinfo {author}
  {\bibfnamefont {Z.-S.}\ \bibnamefont {Yuan}}, \ and\ \bibinfo {author}
  {\bibfnamefont {J.-W.}\ \bibnamefont {Pan}},\ }\href {\doibase
  10.1103/PhysRevA.96.011602} {\bibfield  {journal} {\bibinfo  {journal} {Phys.
  Rev. A}\ }\textbf {\bibinfo {volume} {96}},\ \bibinfo {pages} {011602}
  (\bibinfo {year} {2017})}\BibitemShut {NoStop}%
\bibitem [{\citenamefont {Lewenstein}\ \emph {et~al.}(2004)\citenamefont
  {Lewenstein}, \citenamefont {Santos}, \citenamefont {Baranov},\ and\
  \citenamefont {Fehrmann}}]{santosfermibose}%
  \BibitemOpen
  \bibfield  {author} {\bibinfo {author} {\bibfnamefont {M.}~\bibnamefont
  {Lewenstein}}, \bibinfo {author} {\bibfnamefont {L.}~\bibnamefont {Santos}},
  \bibinfo {author} {\bibfnamefont {M.~A.}\ \bibnamefont {Baranov}}, \ and\
  \bibinfo {author} {\bibfnamefont {H.}~\bibnamefont {Fehrmann}},\ }\href
  {\doibase 10.1103/PhysRevLett.92.050401} {\bibfield  {journal} {\bibinfo
  {journal} {Phys. Rev. Lett.}\ }\textbf {\bibinfo {volume} {92}},\ \bibinfo
  {pages} {050401} (\bibinfo {year} {2004})}\BibitemShut {NoStop}%
\bibitem [{\citenamefont {Donohue}\ \emph {et~al.}(2001)\citenamefont
  {Donohue}, \citenamefont {Tsuchiizu}, \citenamefont {Giamarchi},\ and\
  \citenamefont {Suzumura}}]{danohue2001}%
  \BibitemOpen
  \bibfield  {author} {\bibinfo {author} {\bibfnamefont {P.}~\bibnamefont
  {Donohue}}, \bibinfo {author} {\bibfnamefont {M.}~\bibnamefont {Tsuchiizu}},
  \bibinfo {author} {\bibfnamefont {T.}~\bibnamefont {Giamarchi}}, \ and\
  \bibinfo {author} {\bibfnamefont {Y.}~\bibnamefont {Suzumura}},\ }\href
  {\doibase 10.1103/PhysRevB.63.045121} {\bibfield  {journal} {\bibinfo
  {journal} {Phys. Rev. B}\ }\textbf {\bibinfo {volume} {63}},\ \bibinfo
  {pages} {045121} (\bibinfo {year} {2001})}\BibitemShut {NoStop}%
\bibitem [{\citenamefont {Feiguin}\ and\ \citenamefont
  {Fisher}(2011)}]{FeiguinFisher2011}%
  \BibitemOpen
  \bibfield  {author} {\bibinfo {author} {\bibfnamefont {A.~E.}\ \bibnamefont
  {Feiguin}}\ and\ \bibinfo {author} {\bibfnamefont {M.~P.~A.}\ \bibnamefont
  {Fisher}},\ }\href {\doibase 10.1103/PhysRevB.83.115104} {\bibfield
  {journal} {\bibinfo  {journal} {Phys. Rev. B}\ }\textbf {\bibinfo {volume}
  {83}},\ \bibinfo {pages} {115104} (\bibinfo {year} {2011})}\BibitemShut
  {NoStop}%
\bibitem [{\citenamefont {Cr\'epin}\ \emph {et~al.}(2011)\citenamefont
  {Cr\'epin}, \citenamefont {Laflorencie}, \citenamefont {Roux},\ and\
  \citenamefont {Simon}}]{Simon2011}%
  \BibitemOpen
  \bibfield  {author} {\bibinfo {author} {\bibfnamefont {F.~m.~c.}\
  \bibnamefont {Cr\'epin}}, \bibinfo {author} {\bibfnamefont {N.}~\bibnamefont
  {Laflorencie}}, \bibinfo {author} {\bibfnamefont {G.}~\bibnamefont {Roux}}, \
  and\ \bibinfo {author} {\bibfnamefont {P.}~\bibnamefont {Simon}},\ }\href
  {\doibase 10.1103/PhysRevB.84.054517} {\bibfield  {journal} {\bibinfo
  {journal} {Phys. Rev. B}\ }\textbf {\bibinfo {volume} {84}},\ \bibinfo
  {pages} {054517} (\bibinfo {year} {2011})}\BibitemShut {NoStop}%
\bibitem [{\citenamefont {Okumura}\ \emph {et~al.}(2011)\citenamefont
  {Okumura}, \citenamefont {Yamada}, \citenamefont {Machida},\ and\
  \citenamefont {Aoki}}]{Okumura2011}%
  \BibitemOpen
  \bibfield  {author} {\bibinfo {author} {\bibfnamefont {M.}~\bibnamefont
  {Okumura}}, \bibinfo {author} {\bibfnamefont {S.}~\bibnamefont {Yamada}},
  \bibinfo {author} {\bibfnamefont {M.}~\bibnamefont {Machida}}, \ and\
  \bibinfo {author} {\bibfnamefont {H.}~\bibnamefont {Aoki}},\ }\href {\doibase
  10.1103/PhysRevA.83.031606} {\bibfield  {journal} {\bibinfo  {journal} {Phys.
  Rev. A}\ }\textbf {\bibinfo {volume} {83}},\ \bibinfo {pages} {031606}
  (\bibinfo {year} {2011})}\BibitemShut {NoStop}%
\bibitem [{\citenamefont {Degli Esposti~Boschi}\ \emph
  {et~al.}(2016)\citenamefont {Degli Esposti~Boschi}, \citenamefont
  {Montorsi},\ and\ \citenamefont {Roncaglia}}]{Marco2016}%
  \BibitemOpen
  \bibfield  {author} {\bibinfo {author} {\bibfnamefont {C.}~\bibnamefont
  {Degli Esposti~Boschi}}, \bibinfo {author} {\bibfnamefont {A.}~\bibnamefont
  {Montorsi}}, \ and\ \bibinfo {author} {\bibfnamefont {M.}~\bibnamefont
  {Roncaglia}},\ }\href {\doibase 10.1103/PhysRevB.94.085119} {\bibfield
  {journal} {\bibinfo  {journal} {Phys. Rev. B}\ }\textbf {\bibinfo {volume}
  {94}},\ \bibinfo {pages} {085119} (\bibinfo {year} {2016})}\BibitemShut
  {NoStop}%
\bibitem [{\citenamefont {Liu}\ \emph {et~al.}(2019)\citenamefont {Liu},
  \citenamefont {Yuan}, \citenamefont {Lin},\ and\ \citenamefont
  {Yan}}]{Liu2019}%
  \BibitemOpen
  \bibfield  {author} {\bibinfo {author} {\bibfnamefont {W.-L.}\ \bibnamefont
  {Liu}}, \bibinfo {author} {\bibfnamefont {T.-Z.}\ \bibnamefont {Yuan}},
  \bibinfo {author} {\bibfnamefont {Z.}~\bibnamefont {Lin}}, \ and\ \bibinfo
  {author} {\bibfnamefont {W.}~\bibnamefont {Yan}},\ }\href {\doibase
  10.1088/1674-1056/28/2/020303} {\bibfield  {journal} {\bibinfo  {journal}
  {Chinese Physics B}\ }\textbf {\bibinfo {volume} {28}},\ \bibinfo {pages}
  {020303} (\bibinfo {year} {2019})}\BibitemShut {NoStop}%
\bibitem [{\citenamefont {Giamarchi}(2003)}]{giamarchi2003quantum}%
  \BibitemOpen
  \bibfield  {author} {\bibinfo {author} {\bibfnamefont {T.}~\bibnamefont
  {Giamarchi}},\ }\href@noop {} {\emph {\bibinfo {title} {Quantum physics in
  one dimension}}},\ Vol.\ \bibinfo {volume} {121}\ (\bibinfo  {publisher}
  {Clarendon press},\ \bibinfo {year} {2003})\BibitemShut {NoStop}%
\bibitem [{\citenamefont {Fabrizio}(1993)}]{fabrizio}%
  \BibitemOpen
  \bibfield  {author} {\bibinfo {author} {\bibfnamefont {M.}~\bibnamefont
  {Fabrizio}},\ }\href {\doibase 10.1103/PhysRevB.48.15838} {\bibfield
  {journal} {\bibinfo  {journal} {Phys. Rev. B}\ }\textbf {\bibinfo {volume}
  {48}},\ \bibinfo {pages} {15838} (\bibinfo {year} {1993})}\BibitemShut
  {NoStop}%
\bibitem [{\citenamefont {Nagaosa}(1995)}]{Nagaosa1995}%
  \BibitemOpen
  \bibfield  {author} {\bibinfo {author} {\bibfnamefont {N.}~\bibnamefont
  {Nagaosa}},\ }\href
  {http://www.sciencedirect.com/science/article/pii/0038109895001344}
  {\bibfield  {journal} {\bibinfo  {journal} {Solid State Communications}\
  }\textbf {\bibinfo {volume} {94}},\ \bibinfo {pages} {495} (\bibinfo {year}
  {1995})}\BibitemShut {NoStop}%
\bibitem [{\citenamefont {Schulz}(1996)}]{schulz1996}%
  \BibitemOpen
  \bibfield  {author} {\bibinfo {author} {\bibfnamefont {H.~J.}\ \bibnamefont
  {Schulz}},\ }\href {\doibase 10.1103/PhysRevB.53.R2959} {\bibfield  {journal}
  {\bibinfo  {journal} {Phys. Rev. B}\ }\textbf {\bibinfo {volume} {53}},\
  \bibinfo {pages} {R2959} (\bibinfo {year} {1996})}\BibitemShut {NoStop}%
\bibitem [{\citenamefont {Balents}\ and\ \citenamefont
  {Fisher}(1996)}]{balents1996}%
  \BibitemOpen
  \bibfield  {author} {\bibinfo {author} {\bibfnamefont {L.}~\bibnamefont
  {Balents}}\ and\ \bibinfo {author} {\bibfnamefont {M.~P.~A.}\ \bibnamefont
  {Fisher}},\ }\href {\doibase 10.1103/PhysRevB.53.12133} {\bibfield  {journal}
  {\bibinfo  {journal} {Phys. Rev. B}\ }\textbf {\bibinfo {volume} {53}},\
  \bibinfo {pages} {12133} (\bibinfo {year} {1996})}\BibitemShut {NoStop}%
\bibitem [{\citenamefont {Noack}\ \emph {et~al.}(1996)\citenamefont {Noack},
  \citenamefont {White},\ and\ \citenamefont {Scalapino}}]{Noack1996}%
  \BibitemOpen
  \bibfield  {author} {\bibinfo {author} {\bibfnamefont {R.~M.}\ \bibnamefont
  {Noack}}, \bibinfo {author} {\bibfnamefont {S.~R.}\ \bibnamefont {White}}, \
  and\ \bibinfo {author} {\bibfnamefont {D.~J.}\ \bibnamefont {Scalapino}},\
  }\href {http://www.sciencedirect.com/science/article/pii/S0921453496005151}
  {\bibfield  {journal} {\bibinfo  {journal} {Physica C: Superconductivity}\
  }\textbf {\bibinfo {volume} {270}},\ \bibinfo {pages} {281} (\bibinfo {year}
  {1996})}\BibitemShut {NoStop}%
\bibitem [{\citenamefont {White}\ \emph {et~al.}(2002)\citenamefont {White},
  \citenamefont {Affleck},\ and\ \citenamefont {Scalapino}}]{white2002}%
  \BibitemOpen
  \bibfield  {author} {\bibinfo {author} {\bibfnamefont {S.~R.}\ \bibnamefont
  {White}}, \bibinfo {author} {\bibfnamefont {I.}~\bibnamefont {Affleck}}, \
  and\ \bibinfo {author} {\bibfnamefont {D.~J.}\ \bibnamefont {Scalapino}},\
  }\href {\doibase 10.1103/PhysRevB.65.165122} {\bibfield  {journal} {\bibinfo
  {journal} {Phys. Rev. B}\ }\textbf {\bibinfo {volume} {65}},\ \bibinfo
  {pages} {165122} (\bibinfo {year} {2002})}\BibitemShut {NoStop}%
\bibitem [{\citenamefont {Donohue}\ and\ \citenamefont
  {Giamarchi}(2001{\natexlab{a}})}]{Giamarchi2001}%
  \BibitemOpen
  \bibfield  {author} {\bibinfo {author} {\bibfnamefont {P.}~\bibnamefont
  {Donohue}}\ and\ \bibinfo {author} {\bibfnamefont {T.}~\bibnamefont
  {Giamarchi}},\ }\href {\doibase 10.1103/PhysRevB.63.180508} {\bibfield
  {journal} {\bibinfo  {journal} {Phys. Rev. B}\ }\textbf {\bibinfo {volume}
  {63}},\ \bibinfo {pages} {180508} (\bibinfo {year}
  {2001}{\natexlab{a}})}\BibitemShut {NoStop}%
\bibitem [{\citenamefont {Luthra}\ \emph {et~al.}(2008)\citenamefont {Luthra},
  \citenamefont {Mishra}, \citenamefont {Pai},\ and\ \citenamefont
  {Das}}]{Tapan2008}%
  \BibitemOpen
  \bibfield  {author} {\bibinfo {author} {\bibfnamefont {M.~S.}\ \bibnamefont
  {Luthra}}, \bibinfo {author} {\bibfnamefont {T.}~\bibnamefont {Mishra}},
  \bibinfo {author} {\bibfnamefont {R.~V.}\ \bibnamefont {Pai}}, \ and\
  \bibinfo {author} {\bibfnamefont {B.~P.}\ \bibnamefont {Das}},\ }\href
  {\doibase 10.1103/PhysRevB.78.165104} {\bibfield  {journal} {\bibinfo
  {journal} {Phys. Rev. B}\ }\textbf {\bibinfo {volume} {78}},\ \bibinfo
  {pages} {165104} (\bibinfo {year} {2008})}\BibitemShut {NoStop}%
\bibitem [{\citenamefont {Singh}\ \emph {et~al.}(2014)\citenamefont {Singh},
  \citenamefont {Mishra}, \citenamefont {Pai},\ and\ \citenamefont
  {Das}}]{singh1}%
  \BibitemOpen
  \bibfield  {author} {\bibinfo {author} {\bibfnamefont {M.}~\bibnamefont
  {Singh}}, \bibinfo {author} {\bibfnamefont {T.}~\bibnamefont {Mishra}},
  \bibinfo {author} {\bibfnamefont {R.~V.}\ \bibnamefont {Pai}}, \ and\
  \bibinfo {author} {\bibfnamefont {B.~P.}\ \bibnamefont {Das}},\ }\href
  {\doibase 10.1103/PhysRevA.90.013625} {\bibfield  {journal} {\bibinfo
  {journal} {Phys. Rev. A}\ }\textbf {\bibinfo {volume} {90}},\ \bibinfo
  {pages} {013625} (\bibinfo {year} {2014})}\BibitemShut {NoStop}%
\bibitem [{\citenamefont {Donohue}\ and\ \citenamefont
  {Giamarchi}(2001{\natexlab{b}})}]{giamarchiladder}%
  \BibitemOpen
  \bibfield  {author} {\bibinfo {author} {\bibfnamefont {P.}~\bibnamefont
  {Donohue}}\ and\ \bibinfo {author} {\bibfnamefont {T.}~\bibnamefont
  {Giamarchi}},\ }\href {\doibase 10.1103/PhysRevB.63.180508} {\bibfield
  {journal} {\bibinfo  {journal} {Phys. Rev. B}\ }\textbf {\bibinfo {volume}
  {63}},\ \bibinfo {pages} {180508} (\bibinfo {year}
  {2001}{\natexlab{b}})}\BibitemShut {NoStop}%
\bibitem [{\citenamefont {Cazalilla}\ \emph {et~al.}(2006)\citenamefont
  {Cazalilla}, \citenamefont {Ho},\ and\ \citenamefont
  {Giamarchi}}]{Cazalilla2006}%
  \BibitemOpen
  \bibfield  {author} {\bibinfo {author} {\bibfnamefont {M.~A.}\ \bibnamefont
  {Cazalilla}}, \bibinfo {author} {\bibfnamefont {A.~F.}\ \bibnamefont {Ho}}, \
  and\ \bibinfo {author} {\bibfnamefont {T.}~\bibnamefont {Giamarchi}},\ }\href
  {\doibase 10.1088/1367-2630/8/8/158} {\ \textbf {\bibinfo {volume} {8}},\
  \bibinfo {pages} {158} (\bibinfo {year} {2006})}\BibitemShut {NoStop}%
\end{thebibliography}%

\end{document}